\documentclass[journal, 11pt, draftcls, onecolumn]{IEEEtran}
\usepackage[top=0.9in, bottom=0.85in, left=1in, right=1in]{geometry}
\usepackage{cite}
\usepackage{footnote}
\usepackage{amsmath,amsfonts,amssymb,graphicx}
\usepackage{graphicx}
\usepackage{ifpdf}
\usepackage{amsmath}
\ifCLASSINFOpdf
\else
\fi

\begin{document}
\bibliographystyle{IEEEtran}  
\setlength\arraycolsep{1.5pt}
\title{Interference-Aware Scheduling for Connectivity in MIMO Ad Hoc Multicast Networks}
\author{Feng Jiang, Jianqi Wang, and A. Lee Swindlehurst\thanks{Copyright (c) 2011 IEEE. Personal use of
this material is permitted. However, permission to use this material for any other
purposes must be obtained from the IEEE by sending a request to
pubs-permissions@ieee.org. This work was supported by the National Science Foundation under grant CCF-0916073.  The authors are with the Department of Electrical Engineering and Computer Science, University of California at Irvine, Irvine, CA, 92697 (Email: \{feng.jiang, swindle\}@uci.edu, jianqi.wang@gmail.com).}}


%


\maketitle

\begin{abstract}
We consider a multicast scenario involving an ad hoc network of
co-channel MIMO nodes in which a source node attempts to share a
streaming message with all nodes in the network via some pre-defined
multi-hop routing tree.  The message is assumed to be broken down into
packets, and the transmission is conducted over multiple frames.  Each
frame is divided into time slots, and each link in the routing tree is
assigned one time slot in which to transmit its current packet.  We
present an algorithm for determining the number of time slots and the
scheduling of the links in these time slots in order to optimize the
connectivity of the network, which we define to be the probability
that all links can achieve the required throughput.  In addition to
time multiplexing, the MIMO nodes also employ beamforming to manage
interference when links are simultaneously active, and the beamformers
are designed with the maximum connectivity metric in mind.  The
effects of outdated channel state information (CSI) are taken into
account in both the scheduling and the beamforming designs.  We also
derive bounds on the network connectivity and sum transmit power in
order to illustrate the impact of interference on network performance.
Our simulation results demonstrate that the choice of the number of
time slots is critical in optimizing network performance, and
illustrate the significant advantage provided by multiple antennas in
improving network connectivity.
\end{abstract}



%
\IEEEpeerreviewmaketitle

\newpage
\section{Introduction}

\subsection{Motivation}

Interference management in co-channel ad hoc networks is a challenging
problem.  Simple time-division multiple access (TDMA)-based designs
are inefficient and usually result in relatively poor performance.
While multiple antennas are useful in enhancing channel gain and
reducing interference, incorporating the extra degrees of freedom
offered by MIMO (multi-input, multi-output) nodes into the design of
the network further complicates matters.  Designs based purely on
spatial-division multiple access (SDMA) are not appropriate for large
networks, since the number of available antennas is usually
insufficient for cancelling all of the co-channel interference.
Consequently, space-time (STDMA) solutions must generally be employed,
in which multiple network links are scheduled into each time slot, and
beamforming techniques are used within each slot to mitigate the
resulting interference.

Prior work that has addressed STDMA scheduling for ad hoc networks has
typically focused on finding solutions that maximize the sum
throughput of the network \cite{Pun:2008,Ma:2009,
Zheng:2010,ElBatt:2007}.  However, such solutions inevitably lead to
links with poor performance and localized network congestion, which
cannot be tolerated in applications where the network must perform
multicast streaming \cite{Cordeiro:2003}.  An alternative is of course
to use techniques that ensure fairness ({\em i.e.}, max-min rate,
proportional-fair scheduling, {\em etc.})
\cite{Somarriba:2007,Yue:2008,Marfia:2008}, but such techniques
typically do not directly address network reliability.  Performance
may be fair, but how likely are the links able to achieve this
performance?

The goal of this paper is to suggest methods for addressing the above
issues using physical (PHY) layer techniques in combination with
interference-aware scheduling.  We introduce a novel definition of
network connectivity that quantifies the probability that all links in
the ad hoc network are able to achieve a certain pre-specified
throughput.  The PHY-layer parameters (beamforming vectors, transmit
powers) and the scheduling decisions are then made to maximize this
connectivity metric, taking into account the interference produced by
links that are simultaneously active.  The scenario we have in mind is
ad hoc multicast network streaming, in which a source node attempts to
transmit a continuous data stream to all other nodes in the ad hoc
network with maximum reliability.  Applications of this problem
include tactical military networks (source is the unit commander,
other nodes are teams or individuals under its direction), sensor
surveillance networks (source is a sensor streaming data from a
detected event, network nodes are monitoring stations attempting to
form a coordinated response), vehicular ad hoc networks (VANETs), etc.
For example, a typical VANET scenario involves a vehicle in the
network that detects an unsafe road condition that must be reported to
all other network nodes.  In such a case, throughput is not the most
important aspect of the network, since the actual message may be
relatively short.  Instead, the reliability of the network in sharing
the message from the source with all the vehicles in the network is
the key to achieving safety \cite{Boukerche:2009, Ng:2011}.  
Mobile sensing with distributed platforms ({\em e.g.,} ground-based
robots or UAVs) is another VANET application where connectivity is
more critical than throughput, at least until a detection occurs.  In
sensing mode, such a network must continually be connected in order to
arrive at consensus regarding the parameter(s) being sensed, or until
one of them detects an object of interest, in which case the network
may reconfigure itself for high-throughput data gathering.

Connectivity is fundamentally different from both throughput and
energy consumption, which are the metrics most commonly used to
quantify the performance of a wireless network.  Connectivity
performance is more relevant for applications where robustness or
reliability is the most critical factor, applications (e.g., such as
in military or emergency response scenarios) where the overriding
concern is ensuring that information is shared with all nodes in the
network at some pre-determined minimum rate.  Throughput and energy
consumption do not address network robustness.  Instead, we argue that
the connectivity of the network, as defined in the paper, is a
reasonable way to measure the network's robustness, and thus we set
about choosing the network parameters to maximize the connectivity
metric.  A solution that achieved a higher throughput or lower energy
consumption than our solution would necessarily have poorer
connectivity and hence less robustness, and thus be a less desirable
operating condition for the type of applications we assume.

Our focus is on how the use of multiple antennas in the network
can lead to dramatic improvements in connectivity.  The additional
spatial degrees of freedom offered by the antennas reduce interference
between the links, they allow a ``denser'' scheduling of users for a
given set of resources, and they reduce the amount of transmit power
(and hence interference to other links) to achieve a given throughput.
The combined gain of these effects can only be observed by considering
the joint optimization of the transmit beamformers, scheduling and
power control.  Unlike individuals with hand-held communication
devices, VANET nodes are typically not power constrained and are large
enough to support the use of multiple antennas.  Thus, vehicular
communication systems are a natural platform on which to consider the
performance of multiple-antenna communications.  Since mobility is a
defining aspect of VANETs, it is important to account for the impact
of the resulting time-varying wireless channels.  Unlike most work in
multiple-antenna ad hoc networks, we explicitly take the time-varying
channels into account and quantify their impact on the reliability
(connectivity) of the network.

\subsection{Background}

This work draws on ideas and techniques that have been studied by
many others, but in different contexts, including connectivity,
beamforming for interference networks, and interference-based
scheduling.  Relevant prior work in these areas is briefly discussed
below.

Investigations of radio network connectivity have been conducted by
researchers over the last several decades.  The original connectivity
paradigm was expressed in terms of the so-called ``geometric disk''
model and percolation theory. In the geometric disk model, two nodes
are assumed to be directly connected if their distance is smaller than
some minimum transmission radius.  This results in a simple binary
description of connectivity ({\em i.e.,} the network either is or is
not connected) but lacks an indication of the quality of information
flow.  Percolation theory revolves around finding node density
conditions under which a given node belongs to an unbounded cluster of
connected nodes \cite{BBF02,QT00}. However, neither of these
approaches is reasonable for realistic networks, where one must
consider the effects of fading and interference.  The impact of fading
on network connectivity has been addressed in
\cite{Miorandi:2008,Ibrahim:2009,Han:2009,Yousefi'zadech:2009}.  Of
particular interest to this paper is \cite{Miorandi:2008}, which
showed that multiple antennas can significantly reduce the node
isolation probability in an interference-free network, and
\cite{Yousefi'zadech:2009}, whose simulation results demonstrated that
multiple antennas can significantly enhance the connectivity of an ad
hoc network, measured as the number of links that meet a given
requirement on the outage capacity, or the symbol error rate of
orthogonal space-time block coding.  Interference aspects of the
problem have been studied in \cite{Rajagopalan:2009}, which
investigated the connectivity of sensor networks with regular
topologies.  The network connectivity was defined as the probability
that a path exits between any two pairs of nodes in the network, and
simulation results illustrated how an increase in node density led to
decreased connectivity due to interference effects.

To reduce the impact of interference in ad hoc networks, interference
graph and coloring techniques have been used in the design of
scheduling or routing algorithms \cite{Jain:2005, Ng:2007,
Behzad:2007}. In \cite{Jain:2005}, a linear programming (LP) method
was proposed for computing lower and upper bounds for the maximum
throughput that can be supported by a multi-hop network. A conflict
(interference) graph was used to find the constraint conditions for
the LP formulation. In \cite{Ng:2007}, the authors proposed the
construction of a link-directional interference graph to account for
the directional traffic over each network link. They investigated a
coloring algorithm with two colors on the interference graph to
schedule transmissions in ad hoc networks employing TDMA or frequency
division multiple access (FDMA). In \cite{Behzad:2007}, active links
in a multi-hop network are scheduled in an STDMA scheme where a frame
is divided into equal length time slots and each time slot can be
allocated to several links simultaneously.  Utilizing the interference
graph, a heuristic algorithm is proposed to minimize the frame length
under a constraint that each link's minimum data rate requirement is
satisfied.  An earlier version of this paper \cite{JiangWS10}
discusses the use of interference graphs together with MIMO for
improved connectivity.

The use of multiple antennas to improve the performance of ad hoc
networks has been a topic of considerable recent research
\cite{Zorzi:2006}.  For example, in \cite{Ma:2008}, the transmitter
for each link uses the principal singular vector of the channel matrix
as the transmit beamformer and each receiver uses MMSE beamforming to
mitigate the inter-link interference, and it was shown that there
exists an optimal number of active links that maximize the network
throughput.  Beyond this value, the network becomes
interference-limited, and performance is degraded.  The results in
\cite{Vaze:2009} show if each link transmits a single data stream
without CSI, while the receiver uses part\-ial-zero-forcing
interference cancellation, the capacity lower bound increases linearly
with the number of antennas.  In \cite{Zheng:2010}, an optimal
scheduling policy is proposed that can maximize the average sum rate
of the MIMO ad hoc network under the constraint that the average data
rate of each link is larger than a certain threshold. In
\cite{Yue:2008}, two optimization problems are considered: one is to
maximize the sum rate under sum power and proportional fairness
constraints, and the other is to minimize the sum transmit power under
a constraint on the minimum data rate of each link. In
\cite{Pun:2008}, several distributed scheduling methods were proposed
for MIMO ad hoc networks. In these methods, the links whose channel
condition satisfy a pre-defined threshold are divided into groups for
simultaneous data transmission to maximize the overall network
throughput.  Also relevant is the recent research on MIMO interference
networks, where techniques based on interference alignment
\cite{Gou:2010} and game theory \cite{NoklebyS08} have been proposed.

\subsection{General Approach and Contributions}

As mentioned earlier, in this paper we consider PHY-layer optimization
and scheduling for an ad hoc MIMO network in multicast streaming mode.
In particular, we assume a source node desires to stream data to all
other nodes in the network via a pre-determined multi-hop routing
table (for example, a minimum spanning tree). Our emphasis will
be on obtaining high reliability and low congestion for the network by
maximizing the network connectivity, which we define as the
probability that all links are able to support the desired
throughput. The contributions of this work are as follows:
\begin{enumerate}
\item We examine the multi-hop multicast streaming problem using a
detailed PHY-layer model, and demonstrate the significant benefit that
multiple antennas, power control and proper scheduling can have on
network robustness.
\item We propose a new definition of ad hoc network connectivity that
approximates the probability that all links in the network can achieve
a certain average throughput.  The metric provides a continuous
measure of connectivity performance that is more descriptive than the
simple binary metrics often used in wireless networks.  In addition,
this metric leads to a more robust solution for fast fading channels
than techniques based on requiring that only the {\em average} rate of
each link be above some threshold \cite{Zheng:2010,Yue:2008}.
\item Since optimal connectivity will require that network links be
simultaneously active, we develop beamforming and power control
algorithms for the MIMO interference channels in each time slot that
maximize the network connectivity metric.
\item Together with the beamforming algorithm, we propose an approach
for STDMA scheduling based on greedy coloring of the interference
graph that finds near-optimal assignments of links to each time slot.
\item Unlike nearly all prior work in this area, both the beamforming
and scheduling algorithms take into account the fact that the channel
state information (CSI) may be outdated when it is used, and we
illustrate the impact of the outdated CSI on connectivity.  This is
particularly important for VANET applications, where the network nodes
are mobile.
\item We illustrate via a number of simulations the dramatic
improvement in connectivity that can be obtained when the network
nodes are equipped with multiple antennas.  By jointly considering the
problems of beamformer design, scheduling and power control, we
observe that the use of multiple antennas provides a
``multiplicative'' benefit'' that exceeds what one would expect from
their use in addressing the problems individually.  Furthermore, our
simulations indicate that adding antennas to the network nodes
actually reduces the relative performance loss due to outdated CSI.
\item We derive analytic expressions for an upper bound on the network 
connectivity and a lower bound on the sum transmit power of the network 
assuming interference-free transmissions and no CSI errors.  These
optimistic bounds provide benchmarks to determine the robustness
of the proposed approach.
\end{enumerate}
We note here that a key limitation of the proposed approach is that
the optimization described above takes place assuming that the
multicast routing tree is pre-determined and remains fixed.  This is
clearly suboptimal since the routing decisions determine the links,
which in turn create the interference environment to be mitigated.  In
principle, a complete cross-layer solution would jointly address the
routing, scheduling and PHY-layer issues all at once, but such a
problem is very complex and remains a topic of future investigation.

The rest of the paper is organized as
follows. Section~\ref{sec:network} describes the assumed network-level
model, introduces the definition of the network connectivity metric
and formulates the general optimization problem to be solved.
Section~\ref{sec:link} then presents the link-level MIMO model, which
includes a description of the time-evolution of the MIMO channels and
the outdated CSI.  Section~\ref{sec:beamform} describes the
max-connectivity beamforming solution, and Section~\ref{sec:sched}
puts everything together in the scheduling algorithm, with an
appropriate power control iteration to reduce the transmit power on
each link to its minimum possible value.  Performance bounds on
connectivity and sum transmit power are derived in
Section~\ref{sec:perf}, and results from a collection of simulation
studies are presented in Section~\ref{sec:sim} to illustrate the
performance of the algorithm.  Section~\ref{sec:conc} then concludes
the paper and summarizes our results. The notation used in the paper 
is summarized in Table \ref{eq:table1}.
\begin{table}\caption{Summary of Notations}\label{eq:table1}
\centering
\begin{tabular}{|l|l|}
\hline 
$N$& Number of nodes in the network\\
$M$& Number of antennas per node\\
$N_L$& Number of links in the network spanning tree\\
$C_{req}$& Data rate requirement for active links\\
$N_s$& Number of time slots in a frame\\
$SINR_k$& The signal-to-interference-plus-noise ratio of the $k$th link\\
$SINR_t$& Threshold for $SINR_k$ to guarantee the data rate requirement\\
$P_{out,k}$& Outage probability for the $k$th link\\
$U$& Network connectivity metric\\
$x_k$& Data symbol transmitted over the $k$th link\\
$\mathbf{t}_k$& Transmit beamformer for the $k$th link\\
$P_k$& Transmit power of the $k$th link\\
$d_{ik}$& Distance between the transmitter of link $i$ and the receiver of link $k$\\
$\alpha$& Path loss exponent\\
$\mathbf{H}_{ik}$& Channel matrix between the transmitter of link $i$ and the receiver of link $k$\\
$\mathbf{H}_{ik}^{e}$& Estimate of the channel between the transmitter of link $i$ and the receiver of link $k$\\
$\sigma_{k,1}$& Largest singular value of $\mathbf{H}_{ik}^{e}$\\
$\mathbf{E}_{ik}$& Channel perturbation for the link between the transmitter of link $i$ and the receiver of link $k$\\
$\gamma$& Channel temporal correlation coefficient\\
$\mathbf{n}_k$& Additive noise vector at receiver of link $k$\\
$\mathbf{w}_k$& Receive beamformer at receiver of link $k$\\
$S_{P,k}$ & Scheduling priority of link $k$\\
\hline
\end{tabular}
\end{table}
\section{Network Model and Connectivity Definition}\label{sec:network}

In this section, we provide a model for the network configuration and
the physical layer channel assumed in this paper.  We also define a
connectivity metric for the network, which is the performance objective
we wish to optimize.

\subsection{Network Configuration and Assumptions}

\begin{figure}
\begin{center}
\includegraphics[height=2.5in, width=3in]{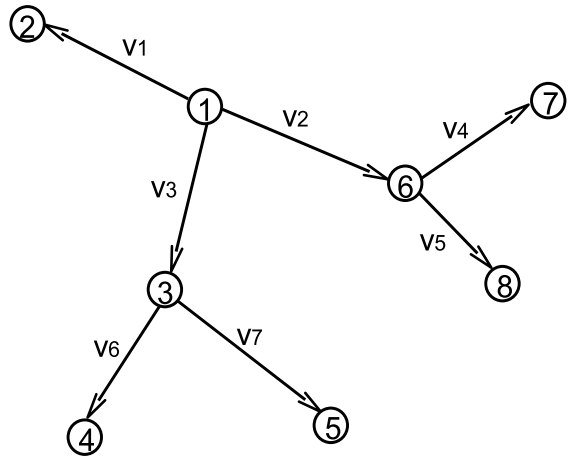}
\caption{Transmission graph based on a multicast network with 8 nodes,
where node 1 is the source node, and $v_i$ is a label used to denote
each link.}\label{f_1}
\vspace{-2em}
\end{center}
\end{figure}

We consider a multi-hop wireless network with a set of $N$ nodes, each
of which is equipped with $M$ antennas (the assumption of an equal
number of antennas is not strictly necessary, but simplifies the
presentation). We assume that a source node wants to share a streaming
message with all other nodes in the network through a pre-defined
routing tree, as depicted for example in Figure~\ref{f_1}.  To
avoid congestion and maintain a constant average data flow from the
source to all nodes, each link must achieve a certain minimum data
rate with high probability.  Performance beyond that achievable with
simple TDMA-based protocols is possible with the availability of
multiple antennas, but co-channel interference must be managed through
appropriate scheduling, power control and transmit/receive
beamforming.  The ability of the network to achieve these goals
depends heavily on the accuracy of the CSI available to the scheduler,
as well as that available to each link of the network.

The scheduling and transmit parameter design are centralized, and are
based on outdated CSI fed back to the scheduler from the individual
links.  Consequently, we avoid the use of spatial multiplexing and
assume that the data on each link is transmitted via a single data
stream using a single transmit beamformer.  The
signal-to-int\-er\-fer\-ence-plus-noise ratio (SINR) of this data
stream determines the rate of the link.  Design of the receive
beamformers and power control is performed by the receivers on each
link, who are assumed to be aware of the statistics of the
interference and the instantaneous effective CSI (channel times
transmit beamformer).  The source message is assumed to be broken down
into packets, and the transmission is conducted over multiple frames.
Each frame is divided into time slots, and each link in the routing
tree is assigned one time slot in which to transmit its current
packet.  The scheduler determines the number of time slots and which
links are active in each time slot, in order to optimize the
connectivity of the network (defined below).  The problem with
dividing the frame into too many slots (emphasizing time rather than
space-time multiplexing of the links) is that, for a fixed frame
duration, the slot length is shorter, thus requiring a higher SINR to
achieve the same throughput over the frame. This is the fundamental
trade-off: fewer slots means more interference, but a lower SINR
requirement; many slots means less interference, but a higher SINR
requirement. We want to find the optimal number of slots for the best
performance.

\subsection{Link Throughput Requirement}

For a given frame, we assume that each of the active links will be
allocated a single time slot, and during this time slot, each
transmitter can occupy the entire available system bandwidth to send
its data packet to the intended receiver. Define $C_{req}$ to be the
minimum rate at which a link can be considered connected, and let the
number of slots in a frame be $N_{s}$. For link $k$ to meet the rate
requirement, its SINR should satisfy the following condition:
\begin{equation}
SINR_{k}\ge SINR_{t},
\end{equation}
where $SINR_{t}=2^{N_{s}C_{req}}-1$. Due to fading and co-channel
interference, the actual $SINR_{k}$ may be smaller than $SINR_t$. A
link is said to experience an outage when the SINR at the receiver is
smaller than the threshold $SINR_{t}$.  The outage probability for link $k$ 
is thus
\begin{equation}
P_{out,k}=\mathrm{Pr}\{SINR_{k}< SINR_t\}.
\end{equation}

\subsection{Definition of Network Connectivity}

We assume that the main goal of the network is \emph{robustness}
rather than throughput.  We want to allocate the resources of the
network (transmit power, beamformers, scheduling) such that the
network remains connected with the highest possible probability, 
where the term ``connected'' implies that each link is able to
communicate at or above the minimum acceptable rate $C_{req}$.  This
in turn requires that an active link must have a certain minimum
SINR. We define the connectivity as the likelihood that all the links
in the network can achieve a SINR that allows transmission at or
above the desired minimum data rate.  Note that this is significantly
different from simply requiring that the {\em average} rate be above
some threshold, which is the approach taken in prior work on fast
fading channels.  

If the network interference has been properly managed so that its
impact on a given link is negligible, then the probability of a
successful transmission on a given link is independent of the other
links, and the connectivity metric can be defined as
\begin{equation}\label{eq:con1}
U=\prod_{k=1}^{N_{L}}(1-P_{out,k}).
\end{equation}
According to this definition, the network connectivity is
equal to the probability that none of the links in the network
experiences an outage during the transmission frame.
Assuming a network of $N_L$ links and a frame with $N_s$ slots, 
$SINR_{k}$ can be expressed as a general function of the 
transmission parameters: 
\begin{equation}\label{eq:formulation}
SINR_{k}=f(\mathbf{t}_{1},\cdots,\mathbf{t}_{N_{L}}, P_{1},\cdots, P_{N_{L}}, N_{s}, \mathcal{S}\;|\mathcal{H}),
\end{equation}
where $\mathbf{t}_{k}\in\mathbb{C}^{M\times1}$ and $\{\mathbf{t}_k| k=1,\cdots,N_L\}$ denotes the link
beamformers, $\{P_1,\cdots,P_{N_L}\}$ is the transmit power allocated
to each link, $\mathbf{H}_{ik}\in \mathbb{C}^{M\times M}$ and $\mathcal{H}=\{\mathbf{H}_{ik} | i,k=1,\cdots,N_{L}\}$
represents the channels between the transmitter of link $i$ and
receiver of link $k$, $\mathcal{S}=\{s_{1},\cdots, s_{N_{L}}\}$
indicates the link scheduling scheme, $s_{k}$ is the slot number
allocated to link $k$, and $s_{k} \in \{1,\cdots,N_{s}\}$.  Based
on~(\ref{eq:formulation}), the connectivity can thus be expressed
as
\begin{equation}\label{eq:ou}
U=\prod_{k=1}^{N_{L}}{\mathbb{E}\{I(SINR_{k})\}}=\prod_{k=1}^{N_{L}}\int{I(SINR_{k}|\mathcal{H})}p(\mathcal{H})d\mathcal{H},
\end{equation}
where $p(\mathcal{H})$ denotes the probability density function (PDF)
of $\mathcal{H}$ and $I(\cdot)$ is the indicator function defined
as
\begin{equation}\label{eq:isinr}
I(SINR_{k})=\left\{\begin{array}{lr}
1 & \textrm{$SINR_{k}\ge SINR_{t}$}\\
0 & \textrm{$SINR_{k}<SINR_{t}$}.
\end{array} \right.
\end{equation}
Under this definition, when the connectivity of the network is
high (near one), the total network throughput would be approximately
bounded below by the number of network links times $C_{req}$.  To
achieve a higher total throughput, $C_{req}$ could be increased, but
this would likely reduce the connectivity.  A desirable trade-off
between connectivity and throughput could conceivably be reached by
appropriately adjusting the minimum link rate $C_{req}$.

The optimization problem associated with maximizing the connectivity
of the network could thus be expressed as
\begin{eqnarray}\label{eq:opt1}
\max\limits_{\{P_{k}\}_{k=1}^{N_L},\{\mathbf{t}_{k}\}_{k=1}^{N_L},N_{s},\mathcal{S}}&&U\\
\mathrm{subject\quad to}&&\|\mathbf{t}_{k}\|=1,\nonumber\\
&&P_{k}\le P_{\max} \; , \; \forall k\in\{1,\cdots,N_L\}\nonumber ,
\end{eqnarray}
where $P_{\max}$ denotes an upper bound on the transmit power.  
Direct optimization of~(\ref{eq:opt1}) is intractable, since one cannot
find a closed-form expression for the probability of~(\ref{eq:isinr}) in 
terms of the parameters of interest.  Instead, we choose to solve the
following related optimization problem:
\begin{eqnarray}\label{eq:opt3}
\max\limits_{\{P_{k}\}_{k=1}^{N_L},\{\mathbf{t}_{k}\}_{k=1}^{N_L},N_{s},\mathcal{S}}&&\sum_{k=1}^{N_{L}}I\big(SINR_{k}|{\mathcal{H}}\big)\\
\mathrm{subject\quad to}&&\|\mathbf{t}_{k}\|=1,\nonumber\\
&&P_{k}\le P_{\max} \; , \; \forall k\in\{1,\cdots,N_L\}\; . \nonumber 
\end{eqnarray}
The relationship between this simplified problem and the original
optimization is analogous to the relationship between the geometric
and arithmetic means.  Instead of maximizing the product of the
link connection probabilities, we maximize its sum.

We divide the solution of~(\ref{eq:opt3}) into two sub-problems: (1)
scheduling and (2) beamformer design and power allocation. We solve
sub-problem (2) for different results of sub-problem (1) to determine
which scheduling result is best.  The transmit beamforming problem
is discussed in the next section, and the scheduling algorithm is 
described in Section~\ref{sec:sched}.

\section{Interference Channel Model with Outdated CSI}\label{sec:link}

\subsection{Interference Channel Model}

Assume without loss of generality that links $1,\cdots,N_k$ are
simultaneously active with link $k$ during a given time slot ($k >
N_k$).  The signal at link $k$'s receiver can be expressed as
\begin{equation}
\mathbf{y}_k=\sqrt{\frac{P_k}{d_{kk}^{\alpha}}}\mathbf{H}_{kk}\mathbf{t}_{k}x_{k}+\sum_{i=1}^{N_k}\sqrt{\frac{P_i}{d_{ik}^{\alpha}}}\mathbf{H}_{ik}\mathbf{t}_{i}x_{i}+\mathbf{n}_k,
\end{equation}
where the transmitted symbol $x_{k}$ is a unit-magnitude data symbol,
$d_{ik}$ is the distance between the transmitter of link $i$ and
receiver of link $k$, $\alpha$ is the path loss exponent, and
$\mathbf{n}_k \in\mathbb{C}^{M\times1}$ is an additive, spatially
white noise vector with covariance given by
$\mathbb{E}\{\mathbf{n}_k\mathbf{n}_k^H\}=\sigma^2\mathbf{I}$ and $\mathbf{I}$ is the $M\times M$ identity matrix.

Assuming receiver $k$ employs a beamformer $\mathbf{w}_k\in\mathbb{C}^{M\times 1}$, the SINR
for link $k$ is given by 
\begin{equation}
SINR_k=\frac{\mathbb{E}\bigg\{\Big|\sqrt{\frac{P_k}{d_{kk}^{\alpha}}}\mathbf{w}_k^H\mathbf{H}_{kk}\mathbf{t}_{k}x_{k}\Big|^2\bigg\}}{\mathbb{E}\bigg\{\Big|\sum_{i=1}^{N_k}\sqrt{\frac{P_i}{d_{ik}^{\alpha}}}\mathbf{w}_k^H\mathbf{H}_{ik}\mathbf{t}_{i}x_{i}+\mathbf{w}_k^H\mathbf{n}_{k}\Big|^2\bigg\}}=\frac{P_k}{d_{kk}^{\alpha}}\frac{\big|\mathbf{w}_k^H\mathbf{H}_{kk}\mathbf{t}_{k}\big|^2}{\mathbf{w}_k^{H}\mathbf{Q}_k\mathbf{w}_k},
\end{equation}
where $\mathbf{Q}_k=\sum_{i=1}^{N_k}\frac{P_i}{d_{ik}^\alpha}\mathbf{H}_{ik}\mathbf{t}_{i}(\mathbf{H}_{ik}\mathbf{t}_{i})^H+\sigma^2\mathbf{I}$. Assuming
the receiver knows the covariance matrix
$\mathbf{Q}_{k}$ and the channel vector
$\mathbf{H}_{kk}\mathbf{t}_{k}$, the optimal $\mathbf{w}_k$ that
maximizes $SINR_k$ is given by \cite{Winters:1984}:
\begin{equation}\label{eq:sinr}
\mathbf{w}_k=\mathbf{Q}_{k}^{-1}\mathbf{H}_{kk}\mathbf{t}_{k},
\end{equation}
and the resulting SINR for link $k$ can be expressed as
\begin{equation}\label{eq:sinra}
SINR_{k}=\frac{P_k}{d_{kk}^\alpha}\mathbf{t}_{k}^H\mathbf{H}_{kk}^{H}\mathbf{Q}_k^{-1}\mathbf{H}_{kk}\mathbf{t}_{k}.
\end{equation}


\subsection{Outdated CSI}\label{subsec:csi}

The CSI at the scheduler will be outdated due to the time required for
this information to be fed back from the network nodes.  To quantify the
CSI uncertainty due to the feedback delay, we adopt a first order
Markov model to characterize the time variation of the channel \cite{Zhou:2005}
\begin{equation}\label{eq:outch}
\mathbf{H}_{ik}=\sqrt{1-\gamma}\mathbf{H}_{ik}^{e}+\sqrt{\gamma}
\mathbf{E}_{ik},\quad i,\;k=1,\ldots,N_{L}
\end{equation}
where $\mathbf{H}_{ik}$ denotes the channel matrix during the data
transmission period, $\mathbf{H}_{ik}^{e}$
represents the channel feedback, and
$\mathbf{E}_{ik}\in\mathbb{C}^{M\times M}$ is a perturbation
matrix. The elements of $\mathbf{H}_{ik}^{e}$ and $\mathbf{E}_{ik}$ are 
assumed to be i.i.d complex Gaussian random variables with distribution
$\mathcal{CN}(0,1)$.  The coefficient $\gamma$ is used to determine the
level of uncertainty in the CSI at the scheduler.  In effect, under 
this model the scheduler assumes a Gaussian distribution for $\mathbf{H}_{ik}$,
with mean $\sqrt{1-\gamma} \mathbf{H}^e_{ik}$ and independent entries with
variance $\gamma$.  

Substituting (\ref{eq:outch}) into (\ref{eq:sinra}), the SINR at the
receiver of link $k$ can be expressed as
\begin{equation}
SINR_{k}=\frac{P_{k}}{d_{kk}^{\alpha}}\mathbf{t}_{k}^{H}\left(\sqrt{1-\gamma}\mathbf{H}_{kk}^{e}+\sqrt{\gamma}\mathbf{E}_{kk}\right)^{H}\mathbf{Q}_{k}^{-1}\left(\sqrt{1-\gamma}\mathbf{H}_{kk}^{e}+\sqrt{\gamma}\mathbf{E}_{kk}\right)\mathbf{t}_{k},
\end{equation}
where
$\mathbf{Q}_{k}\!=\!\sum_{i=1}^{N_{k}}\frac{P_{i}}{d_{ik}^{\alpha}}\left(\sqrt{1-\gamma}\mathbf{H}_{ik}^{e}+\sqrt{\gamma}\mathbf{E}_{ik}\right)\mathbf{t}_{i}\mathbf{t}_{i}^{H}\left(\sqrt{1-\gamma}\mathbf{H}_{ik}^{e}+\sqrt{\gamma}\mathbf{E}_{ik}\right)^{H}+\sigma^2\mathbf{I}.$
It is observed that $SINR_{k}$ is a function of the channel
set $\bar{\mathcal{H}}=\{\mathbf{H}_{ik}^{e}|i,k=1,\ldots,N_{L}\}$ and
the channel perturbation set
$\mathcal{E}\!=\!\{\mathbf{E}_{ik}|i,k\!=\!1,\ldots,N_{L}\}$. Given the
channel set $\bar{\mathcal{H}}$, $SINR_{k}$ is a random variable, the
distribution of which depends on the elements in $\mathcal{E}\!$. The
conditional expectation of $\!SINR_{k}\!$ with respect to $\mathcal{E}\!$ is
given by
\begin{eqnarray}{\label{eq:q}}
\mathbb{E}\{SINR_{k}|\bar{\mathcal{H}}\}&=&\mathbb{E}\bigg\{\frac{P_{k}}{d_{kk}^{\alpha}}\mathbf{t}_{k}^{H}\left(\sqrt{1-\gamma}\mathbf{H}_{kk}^{e}+\sqrt{\gamma}\mathbf{E}_{kk}\right)^{H}\mathbf{Q}_{k}^{-1}\left(\sqrt{1-\gamma}\mathbf{H}_{kk}^{e}+\sqrt{\gamma}\mathbf{E}_{kk}\right)\mathbf{t}_{k}\bigg\}\\
&\overset{a}{=}&\mathbb{E}\bigg\{\frac{P_{k}}{d_{kk}^{\alpha}}\mathbf{t}_{k}^{H}\left(\sqrt{1-\gamma}\mathbf{H}_{kk}^{e}+\sqrt{\gamma}\mathbf{E}_{kk}\right)^{H}\mathbb{E}\Big\{\mathbf{Q}_{k}^{-1}\Big\}\left(\sqrt{1-\gamma}\mathbf{H}_{kk}^{e}+\sqrt{\gamma}\mathbf{E}_{kk}\right)\mathbf{t}_{k}\bigg\}\nonumber\\
&=&\frac{P_{k}}{d_{kk}^{\alpha}}\bigg((1-\gamma)\mathbf{t}_{k}^{H}\mathbf{H}_{kk}^{eH}\mathbb{E}\Big\{\mathbf{Q}_{k}^{-1}\Big\}\mathbf{H}_{kk}^{e}\mathbf{t}_{k}+\gamma \mathrm{tr}\Big(\mathbb{E}\Big\{\mathbf{Q}_{k}^{-1}\Big\}\Big)\bigg)\nonumber, 
\end{eqnarray}
where step (a) is due to the fact that perturbation matrices $\mathbf{E}_{ik} (i,k=1,\ldots,N_{L})$ are independent of each other and $\mathrm{tr}(\cdot)$ denotes the trace operator. Note that the use of the expected value in~(\ref{eq:q}) is due
to the fact that the CSI in $\mathcal{H}$ may not be precisely known.
According to (\ref{eq:outch}), the
channel may have changed from the time it was reported since the
network nodes may be mobile. If precise CSI is available, then the
expectation can be dropped and the instantaneous value of $\mathcal{H}$
can be used instead. Calculation of the term $\mathbb{E}\Big\{\mathbf{Q}_{k}^{-1}\Big\}$ is
very complicated, so instead we use the following lower bound based on
Jensen's inequality\cite[Lemma 4]{Zhang:2008}:
\begin{eqnarray}{\label{eq:approx}}
\mathbb{E}\Big\{\mathbf{Q}_{k}^{-1}\Big\}&\succeq&\mathbb{E}\Big\{\mathbf{Q}_{k}\Big\}^{-1}\nonumber\\
&=&\mathbb{E}\left\{(1-\gamma)\sum_{i=1}^{N_{k}}\frac{P_{i}}{d_{ik}^{\alpha}}\mathbf{H}_{ik}^{e}\mathbf{t}_{i}\mathbf{t}_{i}^{H}\mathbf{H}_{ik}^{eH}+\sqrt{\gamma(1-\gamma)}\sum_{i=1}^{N_{k}}\frac{P_{i}}{d_{ik}^{\alpha}}\Big(\mathbf{H}_{ik}^{e}\mathbf{t}_{i}\mathbf{t}_{i}^{H}\mathbf{E}_{ik}^{H}+\mathbf{E}_{ik}\mathbf{t}_{i}\mathbf{t}_{i}^{H}\mathbf{H}_{ik}^{eH}\Big)\right.\nonumber\\
&&\left.+\;\gamma\bigg\{\sum_{i=1}^{N_{k}}\frac{P_{i}}{d_{ik}^{\alpha}}\mathbf{E}_{ik}\mathbf{t}_{i}\mathbf{t}_{i}^{H}\mathbf{E}_{ik}^{H}\bigg\}+\sigma^2\mathbf{I}\right\}^{-1}\nonumber\\
&=&\left((1-\gamma)\sum_{i=1}^{N_{k}}\frac{P_{i}}{d_{ik}^{\alpha}}\mathbf{H}_{ik}^{e}\mathbf{t}_{i}\mathbf{t}_{i}^{H}\mathbf{H}_{ik}^{eH}+\sqrt{\gamma(1-\gamma)}\sum_{i=1}^{N_{k}}\frac{P_{i}}{d_{ik}^{\alpha}}\mathbb{E}\left\{\mathbf{H}_{ik}^{e}\mathbf{t}_{i}\mathbf{t}_{i}^{H}\mathbf{E}_{ik}^{H}+\mathbf{E}_{ik}\mathbf{t}_{i}\mathbf{t}_{i}^{H}\mathbf{H}_{ik}^{eH}\right\}\right.\nonumber\\
&&\left.+\;\gamma\sum_{i=1}^{N_{k}}\frac{P_{i}}{d_{ik}^{\alpha}}\mathbb{E}\left\{\mathbf{E}_{ik}\mathbf{t}_{i}\mathbf{t}_{i}^{H}\mathbf{E}_{ik}^{H}\right\}+\sigma^2\mathbf{I}\right)^{-1}\nonumber\\
&=&\left(\bigg(\gamma\sum_{i=1}^{N_{k}}\frac{P_{i}}{d_{ik}^{\alpha}}+\sigma^2\bigg)\mathbf{I}+(1-\gamma)\sum_{i=1}^{N_{k}}\frac{P_{i}}{d_{ik}^{\alpha}}\mathbf{H}_{ik}^{e}\mathbf{t}_{i}\mathbf{t}_{i}^{H}\mathbf{H}_{ik}^{eH}\right)^{-1},
\end{eqnarray}
where $\mathbf{A}\succeq\mathbf{B}$ denotes that
$\mathbf{A}-\mathbf{B}$ is a positive semidefinite matrix. In the
above calculation, we use the fact that
$\mathbf{E}_{ik}\mathbf{t}_{i}\in\mathbb{C}^{M\times1}$ is a complex
Gaussian random vector with distribution
$\mathbf{E}_{ik}\mathbf{t}_{i}\thicksim\mathcal{CN}(0,\mathbf{I})$.
Substituting (\ref{eq:approx}) into (\ref{eq:q}), the conditional
expectation of the $SINR_k$ is lower bounded by
\begin{eqnarray}\label{eq:lowb}
\mathbb{E}\{SINR_{k}|\bar{\mathcal{H}}\}&\ge&\frac{P_{k}}{d_{kk}^{\alpha}}(1-\gamma)\mathbf{t}_{k}^{H}\mathbf{H}_{kk}^{eH}\!\!\left(\bigg(\gamma\sum_{i=1}^{N_{k}}\frac{P_{i}}{d_{ik}^{\alpha}}+\sigma^2\bigg)\mathbf{I}+(1-\gamma)\sum_{i=1}^{N_{k}}\frac{P_{i}}{d_{ik}^{\alpha}}\mathbf{H}_{ik}^{e}\mathbf{t}_{i}\mathbf{t}_{i}^{H}\mathbf{H}_{ik}^{eH}\right)^{\!-1}\!\!\mathbf{H}_{kk}^{e}\mathbf{t}_{k}\nonumber \\
&&+ \;\gamma \mathrm{tr}\left(\left(\bigg(\gamma\sum_{i=1}^{N_{k}}\frac{P_{i}}{d_{ik}^{\alpha}}+\sigma^2\bigg)\mathbf{I}+(1-\gamma)\sum_{i=1}^{N_{k}}\frac{P_{i}}{d_{ik}^{\alpha}}\mathbf{H}_{ik}^{e}\mathbf{t}_{i}\mathbf{t}_{i}^{H}\mathbf{H}_{ik}^{eH}\right)^{-1}\right).
\end{eqnarray}


\section{Transmit Beamforming for Connectivity}\label{sec:beamform}

In this section we introduce how the transmit beamformer and 
power are calculated for a given link schedule. Consider a scenario in
which $K$ links are transmitting simultaneously. In this case, the
number of links that can meet the desired rate requirement depends on the
beamformer and transmit power that each link adopts. Define
$\mathbb{E}_{l}\{SINR_{k}|\bar{\mathcal{H}}\}$ as the right hand side
of (\ref{eq:lowb}).  The problem of finding the optimal
beamformer and transmit power based on the outdated CSI
$\bar{\mathcal{H}}$ can be stated as:
\begin{eqnarray}\label{eq:opt4}
\max\limits_{\{P_{k}\}_{k=1}^{K}, \{\mathbf{t}_{k}\}_{k=1}^{K}} && \sum_{k=1}^{K} I\big(\mathbb{E}_{l}\{SINR_{k}|\bar{\mathcal{H}}\}\big)\\
\mathrm{subject\; to}&&\|\mathbf{t}_{k}\|=1,\nonumber\\
                     &&P_{k}\le P_{\max}\nonumber \; , \; \forall k\in\{1,\cdots,K\}\nonumber.                   
\end{eqnarray}
The indicator function in~(\ref{eq:opt4}) is not continuous
and thus the problem is difficult to solve with standard optimization
algorithms. Using the following sigmoid approximation \cite{Wang:2007}:
\begin{equation}\label{eq:approx2}
\tilde{I}\big(\mathbb{E}_{l}\{SINR_{k}|\bar{\mathcal{H}}\}\big)\approx\frac{1}{1+e^{-\beta(\mathbb{E}_{l}\{SINR_{k}|\bar{\mathcal{H}}\}-SINR_{t})}} \; ,
\end{equation}
where $\beta$ is the approximation parameter, the problem can be
converted to finding the maximum value of a constrained continuous
nonlinear multivariable function.  Replacing
$I\big(\mathbb{E}_{l}\{SINR_{k}|\bar{\mathcal{H}}\}\big)$ with
$\tilde{I}\big(\mathbb{E}_{l}\{SINR_{k}|\bar{\mathcal{H}}\}\big)$, the
optimization problem in~(\ref{eq:opt4}) can be approximated as
\begin{eqnarray}\label{eq:opt5}
\max\limits_{\{P_{k}\}_{k=1}^{K}, \{\mathbf{t}_{k}\}_{k=1}^{K}} && \sum_{k=1}^{K} \frac{1}{1+e^{-\beta(\mathbb{E}_{l}\{SINR_{k}|\bar{\mathcal{H}}\}-SINR_{t})}}\\
\mathrm{subject\; to}&&\|\mathbf{t}_{k}\|=1,\nonumber\\
                     &&P_{k}\le P_{\max}\nonumber \; , \; \forall k\in\{1,\cdots,K\}\nonumber.     
\end{eqnarray}
Note that when $\beta$ is small, the sigmoid function
$\tilde{I}\big(\mathbb{E}_{l}\{SINR_{S,k}|\bar{\mathcal{H}}\}\big)$ is
smooth. As $\beta \rightarrow\infty$, the sigmoid function
approaches the indicator function. Starting with relatively small
values for $\beta$ and then increasing it in several steps, the
solution involving the indicator function can be found. For each fixed
value of $\beta$, the problem in (\ref{eq:opt5}) can be solved
numerically using, for example, the active-set method.
The algorithm can be initialized with arbitrary power
allocations, and by setting the transmit beamformer $\mathbf{t}_{k}$ equal
to the principal singular vector of $\mathbf{H}_{kk}^{e}$. 

\section{Scheduling for Maximum Connectivity}\label{sec:sched}

In this section, we propose a scheduling algorithm that maximizes the
connectivity metric defined earlier using the concept of coloring from
graph theory\cite{Ng:2007, Behzad:2007}. The algorithm is based on
the use of the interference and collision graph (ICG) of the network\cite{Zheng:2010, Ng:2007, Behzad:2007, Jain:2005}. 

\subsection{The Interference Graph and Greedy Coloring}

We use an ICG to model the relationships between the active links. In
the ICG, each vertex represents a directional link in the transmission
graph. We define two types of neighbors in the ICG. The first are {\em
interfering} neighbors, which represent links that could be
simultaneously active and hence interfere with one another, and the
second are {\em colliding} neighbors, which represent links that
cannot be active at the same time.  In our application, colliding
links include those that share the same transmitter, or those where
a transmitter in one is a receiver in the other.  All links that are
not colliding are considered to be interfering, although the amount of
interference between two given links could be low if they are far
apart.  In the paragraphs below, we more precisely define the ICG and
concepts related to our scheduling algorithm.

\begin{figure}
\begin{center}
\includegraphics[height=2.6in, width=2.78in]{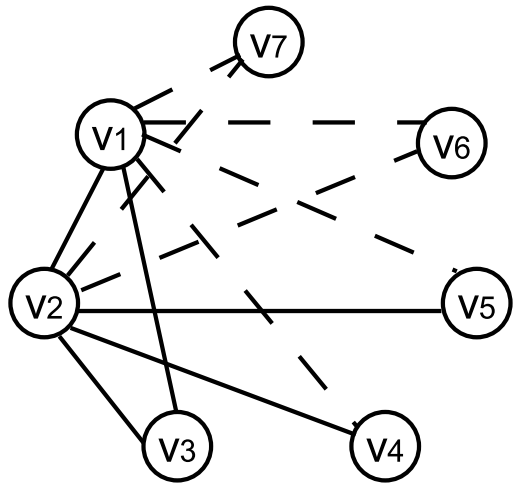}
\caption{Partial ICG for the network of
Fig.~\ref{f_1}, where only edges corresponding to $v_1$ and
$v_2$ are shown.  Colliding links are connected with solid edges,
interfering links with dashed edges.}\label{f_2}
\end{center}
\end{figure}

\emph{Interference and collision graph}: The interference and
collision graph $G_{I}$ can be defined based on the transmission graph
$G_{T}$. A given link $k$ in $G_{T}$ is represented by a vertex
$v_{k}$ in $G_{I}$.  Suppose for link $k$ that node $t_k$ is the
transmitter and $r_k$ is the receiver, and suppose for link $l$ that
$t_l$ is the transmitter and $r_l$ is the receiver.  Links $v_k$ and
$v_l$ are colliding links if any of the following are true: $t_k=t_l,
t_k=r_l$ or $t_l=r_k$ (note that for the multicast tree we assume,
$r_k=r_l$ will never occur). An edge between two vertices in $G_I$
represents that the two corresponding links are colliding links and they could not be assigned to the same time slot; if two
vertices in $G_I$ do not share an edge, they represent interfering links
which can be assigned to the same time slot, provided that the resulting interference could be managed.
As an illustration,
Fig.~\ref{f_2} represents a partial ICG for the network of
Fig.~\ref{f_1}, where the colliding links and interfering
links are connected with solid and dashed edges, respectively (for the
sake of clarity, only edges associated with $v_1$ and $v_2$ are
plotted; the remainder of the edges can be generated in a similar
fashion).  For example, links $v_{2}$ and $v_{1}$ share the same
transmitter node 1, so they are colliding links in Fig.~\ref{f_2};
the receiver of link $v_2$ is the same as the transmitter of link
$v_4$, so $v_{2}$ and $v_{4}$ are also colliding links.

\emph{Coloring}: In our application, ``coloring'' refers to the
process of assigning time slots to the network links, or equivalently,
to the nodes in the interference graph.  Given a set of colors in the
discrete set $\mathcal{C}$ (colors can be considered as distinct
non-negative integers), a coloring of the graph $G$ is an assignment
of the elements (or colors) in $\mathcal{C}$ to the vertices of $G$,
one color for each vertex, such that no adjacent vertices occupy the
same color. A greedy coloring enumerates the vertices in a specific
order $v_1,\dots, v_n$ and assigns $v_k$ to the smallest color that is
not occupied by the neighbors of $v_k$ among $v_1,\dots, v_{k-1}$. The
vertices can be ordered according to their edge degree, which is the
number of edges incident to the vertex \cite[chap. 5]{Diestel:2005}.
To apply coloring to the ICG, we need to define an order for the
vertices. Before we proceed to the scheduling order, some related
definitions are necessary.

\emph{Scheduling freedom}: For a link $k$, the scheduling freedom
$F_{k}$ is the number of available colors that can be allocated to
this link. The higher the value of $F_{k}$, the higher the possibility
that link $k$ will be allocated to a ``good'' color (one that leads to
low interference).

\emph{Collision degree}: Given a link $k$, the collision degree
$C_{D,k}$ is the number of its colliding neighbors in the ICG.

\emph{Constraint and free color sets}: For link $k$, the interfering
color set $\mathcal{D}_{I, k}$ includes colors occupied by the
neighbors that interfere with $v_{k}$. The unavailable color set
$\mathcal{D}_{U,k}$ includes the colors occupied by the neighbors that
collide with $v_{k}$.  The free color set is defined as
$\mathcal{D}_{F,k}=\mathcal{C}-(\mathcal{D}_{I,k}\bigcup\mathcal{D}_{U,k})$,
and corresponds to the set of colors that could be assigned to link
$k$ without causing any interference (or collisions).  The constraint
color set is defined as
$\mathcal{D}_{C,k}=\mathcal{D}_{I,k}-(\mathcal{D}_{I,k}\bigcap\mathcal{D}_{U,k})$.
The colors in this set can also possibly be assigned to link $k$, but
with some additional interference that would have to be mitigated through
beamforming.

\emph{Scheduling priority}: The scheduling order is determined using the
largest singular values $\sigma_{k,1}$ of the channel matrices
$\mathbf{H}_{kk}^e$.  The higher the value of $C_{D,k}$ or the smaller
the channel gain $g_{k}=\sigma_{k,1}^2/d_{kk}^{\alpha}$, the more
likely it is that link $k$ will be affected by interference.  Such a
link will have fewer colors it could be assigned to, and hence a
smaller value of $F_k$.  To increase the likelihood that links with low
scheduling freedom can be allocated a good color, the scheduling
priority of link $k$ is defined as
\begin{equation}
S_{P,k}=C_{D,k}\cdot W+\frac{1}{g_{k}},
\end{equation}
where $W$ is a constant larger than $\max_{k}\frac{1}{g_{k}}$. 

\subsection{Scheduling Algorithm for Connectivity}

Based on the above definitions, we propose here a scheduling algorithm
for optimizing connectivity.  The algorithm assumes a particular value
for the number of slots $N_s$, and is repeated until a value of $N_s$
is found that maximizes the connectivity metric. The minimum possible
value for $N_s$ is the maximum collision degree $\max_k C_{D,k}$ over
all vertices in the ICG. The algorithm
begins by ordering the links according to their scheduling priority,
and then assigns a color to them one-by-one, from highest to lowest priority.
Consider the link at position $m$ in the priority ordering.  If link $m$
can be added to a color that already has had other links assigned to it, 
and if the beamformers and power levels for these links can be adjusted to
accommodate link $m$ without causing any of them to drop below
$SINR_t$, then link $m$ is added to this color.  If there are multiple
colors for which this is true, it is added to the color that requires
the smallest increase in transmit power to accommodate it.  If the
addition of link $m$ to any of these colors causes one of the links
(including possibly link $m$) to drop below the SINR threshold, and if
there exist free colors that have not had any links assigned to them,
link $m$ is assigned to one of the free colors.  If no free colors exist,
then link $m$ is assigned to the color that caused the smallest number
of links to drop below the threshold (and with the smallest increase
in power, in case of a tie).  In this latter case, it is hoped that
the power control algorithm described in the next section will reduce
the interference sufficiently so that all links assigned to the color
will end up being active.  A more precise mathematical description of
the algorithm is given below.

\begin{enumerate}

\item Let $\mathbf{v}_k$ be the right singular vector of
$\mathbf{H}_{kk}^e$ corresponding to the largest singular value, and
let $P_k=P_{\max}$ be the initial transmit power allocated to link $k$.
Assuming a value for $N_s$, initialize the active link set
$\mathcal{A}=\{k \; | \; P_{\max}g_{k}\ge SINR_t, k=1,\dots, N_{L}\}$.
Links that do not qualify for $\mathcal{A}$ cannot meet the desired
target rate for the given value of $N_s$.  Initialize the transmit
beamformers as $\mathcal{T}=\{\mathbf{v}_{k}|k\in\mathcal{A}\}$, the
transmit powers as $\mathcal{P}=\{P_{k}=P_{\max}|k\in\mathcal{A}\}$
and the color set $\mathcal{C}=\{1,2,\cdots,N_{s}\}$.  Let
$\mathcal{P}_{C}=\{P_{C,1},\dots,P_{C,N_{s}}\}$ denote the sum
transmit power of the active links in each color, and let
$\mathcal{N}_{C}=\{N_{C,1},\dots,N_{C,N_{s}}\}$ represent the number
of links that are unable to meet the the target $SINR_t$ for each color.
Initialize these sets to contain all zeros.  The initial schedule
$\mathcal{S}=\{s_{k}|k\in\mathcal{A}\}$ is also set to zeros.
Construct the ICG based on the relationship between the active
links. Compute the scheduling priority $S_{P,k}$ of the vertices
$v_{k}$ for $k\in\mathcal{A}$ .

\item Select the link with the highest scheduling priority: $m=\arg
\max_{k\in\mathcal{A}} S_{P,k}$, and construct the free color set
$\mathcal{D}_{F,m}$ and the constraint color set $\mathcal{D}_{C,m}$ for
link $m$.
\begin{itemize}
\setlength{\parindent}{2em}

\item[] If $\mathcal{D}_{C,m}=\phi$ and $D_{F,m}\neq\phi$

Assign link $m$ to color $j=\min_{i\in \mathcal{D}_{F,m}} i$, set
$s_{m}=j$, $P_{m}=P_{C,j}=\frac{SINR_{t}}{g_{m}}$, and

skip to step 5.

\item[] else if $\mathcal{D}_{C,m}=\phi$ and $D_{F,m}=\phi$

There aren't enough colors to avoid collisions between the active
links, so the

algorithm must be restarted with a larger value for $N_s$.

\item[] else

For $i\in\mathcal{D}_{C,m}$, construct the link set
$\mathcal{L}_{m,i}=\{k \; | \; s_{k}=i \; \mbox{\rm for} \; k \in \mathcal{A}\}$,
which

contains the links currently assigned to color $i$.

\item[] end.
\end{itemize}

\item For each $i\in \mathcal{D}_{C,m}$, assume the links in the set
$\mathcal{L}_{m,i}\bigcup m$ are transmitting simultaneously, and use
the transmit beamforming algorithm of Section~\ref{sec:beamform} to
find the new beamformer and transmit power sets
$\mathcal{T}_{m,i}=\{\mathbf{t}_{m,k}^{i}|k\in\mathcal{L}_{m,i}\bigcup
m\}$, $\mathcal{P}_{m,i}=\{P_{m,k}^{i}|k\in\mathcal{L}_{m,i}\bigcup
m\}$. For link set $\mathcal{L}_{m,i}\bigcup m$, based on
$\mathcal{T}_{m,i}$, $\mathcal{P}_{m,i}$, calculate the expected
number of links that will be unable to meet the SINR threshold with
link $m$ added: $\tilde{N}_{m,i}=\sum_{k\in\mathcal{L}_{m,i}\bigcup
m}{\Big(1-I\big(\mathbb{E}_{l}\{SINR_{k}|\bar{\mathcal{H}}\}\big)\Big)}$,
and calculate the updated sum transmit power due to the addition of
link $m$: $\tilde{P}_{m,i}=\sum_{k\in\mathcal{L}_{m,i}\bigcup
m}P_{m,k}^{i}$.  Set $\Delta N_{m,i}=\tilde{N}_{m,i}-N_{C,i}$ to be
the number of links that will drop below the SINR threshold if link
$m$ is added to color $i$, and define $\Delta
P_{m,i}=\tilde{P}_{m,i}-P_{C,i}$ to be the additional transmit power
required to add link $m$ to color $i$.

\item Find $j^{'}=\arg\min_{i\in\mathcal{D}_{C,m}}{\Delta N_{m,i}}$.
If more than one color corresponds to the minimum $\Delta N_{m,i}$,
select the color with minimum $\Delta P_{m,i}$.
\begin{itemize}
\setlength{\parindent}{2em}
\item[] If $\mathcal{D}_{F,m}=\phi$ or $\Delta N_{m,j'}=0$

Assign link $m$ to color $j^{'}$, set $s_{m}=j^{'}$,
$N_{C,j^{'}}=\tilde{N}_{m,j^{'}}$,
$P_{C,j^{'}}=\tilde{P}_{m,j^{'}}$. Use $\mathcal{T}_{m,j^{'}}$,

$\mathcal{P}_{m,j^{'}}$ to update the components of $\mathcal{T}$ and $\mathcal{P}$ which correspond to
links in the set

$\mathcal{L}_{m,j^{'}}\bigcup m$.

\item[] else if $\mathcal{D}_{F,m}\neq\phi$ and $\Delta N_{m,j}>0$

Assign link $m$ to color $j=\min_{i\in \mathcal{D}_{F,m}}i$ and set
$s_{m}=j$, $P_{m}=P_{C,j}=\frac{SINR_{t}}{g_{m}}$.

\item[] end.
\end{itemize}
\item Set $S_{P,m}=0$, and repeat step 2 until each vertex $v_{k}$,
$k\in\mathcal{A}$, is allocated a color.
\end{enumerate}
Once the scheduling is complete, the active links will transmit data
according to the scheduling result $\mathcal{S}$ using the beamformers
in $\mathcal{T}$ and, at least initially, the transmit powers in
$\mathcal{P}$.  As explained below, the actual transmit power for
each link will be fine-tuned based on feedback from the receivers.


\subsection{Local Power Control for Active Links}

Since the nodes are energy limited, to extend the lifetime of the
network and to reduce the mutual interference caused by the co-channel
links, the transmit power of each link should be minimized under the
constraint of the QoS requirement. Due to the approximation in
(\ref{eq:approx2}), the use of the lower bound in (\ref{eq:lowb}), and
the presence of outdated CSI, the actual $SINR_{k}$ based on
$\mathcal{P}$ and $\mathcal{T}$ will not be exactly equal to the
threshold $SINR_{t}$.  In most instances it will be greater than
$SINR_t$ due to the use of the lower bound in~(\ref{eq:lowb}), but in
some rare cases it can be below the threshold.  To remedy this latter
situation, we reduce the transmit power of any links whose SINR
exceeds the threshold, which reduces co-channel interference and the
transmit power consumed by the network.  For a given time slot
$t\in\mathcal{C}$, the network schedule $\mathcal{S}$ assigns links in
set $\mathcal{L}_{t}=\{k|s_{k}=t,k\in\mathcal{A}\}$ to transmit
simultaneously. The power control algorithm steps through each link
in the time slot, reducing power for the link if its
SINR exceeds the threshold.  A given link may be revisited several
times, since reductions in transmit power for other links reduces the
overall interference, and may allow further reductions in transmit
power for the link.  This process is assumed to repeat a maximum of
$N_a$ times.  If, after all $N_a$ iterations, there are any links
whose SINR is below the threshold, these links are declared to be in
outage, their power is reduced to zero, and an additional $N_b$
iterations are performed to reduce the transmit power even further.
A mathematical description of the power control algorithm for
time slot $t$ is described below.

\emph{Iterative Power Control Algorithm}
\begin{enumerate}
\item Initialize the transmit power and beamformer of link
$k\in\mathcal{L}_t$ with $\mathcal{P}$ and $\mathcal{T}$, set
maximum iteration lengths $N_a$ and $N_b$.

\item Link $k$'s receiver $(k\in\mathcal{L}_t)$ calculates $SINR_{k}$
based on (\ref{eq:sinra}). If $SINR_{k}>SINR_{t}$, the receiver for link $k$
informs the transmitter to reduce
$P_{k}$ to $P_{k}=\frac{P_{k}}{SINR_{k}/SINR_{t}}$.

\item $N_{a}=N_{a}-1$, if $N_{a}>0$, go to step 2. 

\item If $SINR_{k}<SINR_{t}$ for any $k\in\mathcal{L}_t$, set $P_{k}=0$ 
and repeat step 2 for another $N_{b}$ iterations to further reduce the 
transmit power.
\end{enumerate}

\section{Performance Bounds}\label{sec:perf}

In this section, we derive two performance bounds that can be used to
evaluate the limiting behavior of the proposed algorithms.  The first
is an upper bound on the network connectivity metric, and the second
is a lower bound on the average sum transmit power.  Both bounds are
derived under the assumption that the CSI is perfect, and each active
link is free of interference.  When $\gamma$ is small, the
connectivity bound should match the performance of the proposed
approach if the network interference has been properly accounted for.
This is not the case for the bound on transmit power, however, since
the interference mitigation results in beamformers that require excess
power to achieve the rate threshold.  The difference between the
required transmit power and the lower bound represents the price paid
for the enhanced connectivity that results from operating the system
as an interference network.  Due to the complexity of calculating 
the bounds, expressions are derived only for the cases $M=1, 2, 4$.

\subsection{Upper Bound on Network Connectivity}\label{sec:conbound}

The connectivity bound is derived assuming the absence of interference,
and perfect CSI at the scheduler ($\gamma=0$).  Each transmitter uses 
maximum power and selects the principal singular vector of the channel
matrix as its beamformer.  When the receiver is free of co-channel
interference, the resulting signal-to-noise-ratio (SNR) for link $k$ is given by
\begin{equation}
SNR_k=\frac{P_{\max}\sigma_{k,1}^2}{d_{kk}^{\alpha}\sigma^2}.
\end{equation}
If we define $P_{out,k}^{'}=\Pr\{SNR_{k}<SINR_{t}\}$, then the upper
bound on connectivity can be expressed as
\begin{equation}\label{eq:conn}
U_{B}^{M}=\prod_{k=1}^{N_L}(1-P_{out,k}^{'})\ge U.
\end{equation}

The squared singular value $\sigma_{k,1}^2$ corresponds to the largest
eigenvalue of the central Wishart matrix
$\mathbf{H}_{kk}\mathbf{H}_{kk}^H$. Define
$\lambda_{k,1}=\sigma_{k,1}^2$, and note that the cumulative density function
(CDF) of $\lambda_{k,1}$ is given by
\cite[eq. (6)]{Kang:2004}:
\begin{equation}\label{eq:11}
\Pr\{\lambda_{1}\le \lambda\}=\frac{\det(\mathbf{\Phi}(\lambda))}{\left(\prod_{j=1}^{M}{\Gamma^2(j)}\right)},
\end{equation}
where we have dropped the subscript $k$ since the distribution is
assumed to be identical for each link, $\Gamma(\cdot)$ is the gamma
function, $\mathbf{\Phi}(\lambda)$ is an $M\times M$ matrix defined by
$\mathbf{\Phi}(\lambda)_{i,j}=\tilde{\gamma}(i+j-1,\lambda)$, and 
the lower incomplete gamma function $\tilde{\gamma}(n,x)$ has the following 
series expansion:
\begin{equation}
\tilde{\gamma}(n,\lambda)=(n-1)!\left(1-\sum_{k=0}^{n-1}\frac{\lambda^k}{k!}e^{-\lambda}\right).
\end{equation}

For $M=2$, (\ref{eq:11}) reduces to:
\begin{eqnarray}\label{eq:12}
\Pr\{\lambda_{1}\le \lambda\}&=&\frac{\det(\mathbf{\Phi}(\lambda)_{2\times2})}{\prod_{j=1}^{2}{\Gamma^2(j)}}\nonumber\\
&=&\frac{\tilde{\gamma}(1,\lambda)\tilde{\gamma}(3,\lambda)-\tilde{\gamma}^2(2,\lambda)}{\Gamma^2(1)\Gamma^2(2)}\nonumber\\
&=&1-e^{-\lambda}(\lambda^{2}+2)+e^{-2\lambda}.
\end{eqnarray}
Substituting (\ref{eq:12}) into (\ref{eq:conn}), the connectivity upper bound can then be calculated as:
\begin{equation}\label{eq:twoboundc}
U_{B}^{M=2}=e^{-\sum_{k=1}^{N_{L}}\lambda_{\min,k}}\prod_{k=1}^{N_{L}}\left(\lambda_{\min,k}^2-e^{-\lambda_{\min,k}}+2\right) \; ,
\end{equation}
where $\lambda_{\min,k}=\frac{SINR_td_{kk}^{\alpha}\sigma^2}{P_{\max}}$ is the minimum value of the channel gain that can guarantee $SNR_{k}\ge SINR_{t}$.

For $M=4$, after some cumbersome algebra, the CDF of the largest
eigenvalue $\lambda_{1}$ can be expressed as:
\begin{eqnarray}\label{eq:cdffour}
\Pr\{\lambda_{1}\le \lambda\}&=&\frac{\det(\mathbf{\Phi}(\lambda)_{4\times4})}{\prod_{j=1}^{4}{\Gamma^2(j)}}\nonumber\\
&=&1-f_{1}(\lambda)-f_{2}(\lambda)-f_{3}(\lambda)-f_{4}(\lambda),
\end{eqnarray}
where $f_{1}(\lambda)$, $f_{2}(\lambda)$, $f_{3}(\lambda)$, $f_{4}(\lambda)$ are defined in the appendix.
The network connectivity upper bound is then computed as
\begin{eqnarray}\label{eq:fourboundc}
U_{B}^{M=4}&=&\prod_{k=1}^{N_{L}}\Big(f_{1}(\lambda_{\min,k})+f_{2}(\lambda_{\min,k})+f_{3}(\lambda_{\min,k})+f_{4}(\lambda_{\min,k})\Big).
\end{eqnarray}

For the single antenna case ($M=1$), the interference-free $SNR$ at the receiver 
is given by
\begin{equation}
SNR_{k}=\frac{P_{\max}|h_{kk}|^{2}}{d_{kk}^{\alpha}\sigma^2},
\end{equation}
where $|h_{kk}|$ is a Rayleigh random variable. Define $h=|h_{kk}|$,
so that $\Pr\{h\le \lambda\}=1-e^{-\lambda^{2}}$. The probability of a 
successful transmission for the link can be expressed as
\begin{eqnarray}
\Pr\{SNR_{k}\ge SINR_{t}\}&=&1-\Pr\Bigg\{h<\sqrt{\lambda_{\min,k}}\Bigg\}=e^{-\lambda_{\min,k}},
\end{eqnarray}
and thus the network connectivity is given by
\begin{equation}\label{eq:singleboundc}
U_{B}^{M=1}=e^{-\sum_{k=1}^{N_{L}}\lambda_{\min,k}}.
\end{equation}
Comparing (\ref{eq:twoboundc}) and (\ref{eq:fourboundc}) with
(\ref{eq:singleboundc}), it can be observed that
\begin{eqnarray}
R_{2}=\frac{U_{B}^{M=2}}{U_{B}^{M=1}}&=&\prod_{k=1}^{N_{L}}\left(\lambda_{\min,k}^2-e^{-\lambda_{\min,k}}+2\right),\\
R_{4}=\frac{U_{B}^{M=4}}{U_{B}^{M=1}}&=&\prod_{k=1}^{N_{L}}e^{\lambda_{\min,k}}\Big(f_{1}(\lambda_{\min,k})+f_{2}(\lambda_{\min,k})+f_{3}(\lambda_{\min,k})+f_{4}(\lambda_{\min,k})\Big).
\end{eqnarray}
It is easy to verify that both $R_{2}$ and $R_{4}$ are monotonically
increasing functions of $\lambda_{\min,k}$ and larger than 1. $R_{2}$
and $R_{4}$ represent the connectivity gain provided by the use of
multiple antennas.  Since $\lambda_{\min,k}$ is proportional to
$d_{kk}^{\alpha}$, the larger the link distance, the greater the
connectivity gain offered by the use of multiple antennas.

\subsection{Lower Bound on Average Sum Transmit Power}

A lower bound for the average sum transmit power of the network can be
obtained under the assumption that each active link selects its
beamformer as the right singular vector of the channel with largest
singular value, and assuming there is no co-channel interference
between the links.

For link $k$, given $\lambda_{1}$, the power allocation at the transmitter is:
\begin{equation}
P_{k}^{M}=\left\{\begin{array}{ll}
\frac{SINR_{t}d_{kk}^{\alpha}\sigma^2}{\lambda_{1}}, &\textrm{$\lambda_{1}\ge\lambda_{\min,k}$}\\
0, &\textrm{$\lambda_{1}<\lambda_{\min,k}$}
\end{array} \right.
\end{equation}
The average transmit power of link $k$ can be obtained by averaging
$P_{k}^{M}$ over the random variable $\lambda_{1}$. Denote the
probability density function (PDF) of $\lambda_{1}$ as
$f_{\lambda_{1}}(\lambda)$, then $f_{\lambda_{1}}(\lambda)$ can be
explicitly obtained by taking the derivative of (\ref{eq:12}) or
(\ref{eq:cdffour}) with respect to $\lambda$. 

When $M=2$, $f_{\lambda_{1}}^{M=2}(\lambda)$ is given by
\begin{equation}
f_{\lambda_1}^{M=2}(\lambda)=e^{-\lambda}(\lambda^2+2)-2\lambda e^{-\lambda}-2e^{-2\lambda}.
\end{equation}
The average transmit power for link $k$ is calculated as\footnote{In principle,
$\lambda_{\max} \rightarrow \infty$ in this equation, but to numerically evaluate
the integral, we simply choose a large enough value such that the integrand is
essentially zero.}
\begin{eqnarray}
\mathbb{E}\{P_{k}^{M=2}\} & = & \int_{\lambda_{\min,k}}^{\lambda_{\max}}\frac{SINR_{t}d_{kk}^\alpha\sigma^2}{\lambda}f_{\lambda_1}^{M=2}(\lambda)d\lambda\nonumber\\
& = & SINR_{t}d_{kk}^\alpha\sigma^2 \int_{\lambda_{\min,k}}^{\lambda_{\max}}\lambda e^{-\lambda}-2e^{-\lambda}+\frac{2}{\lambda}\big(e^{-\lambda}-e^{-2\lambda}\big)d\lambda \; . \label{eq:power1}
\end{eqnarray}
In (\ref{eq:power1}), the integration of the first two terms can easily 
be found.  To calculate the term
$\int_{}^{}\frac{e^{-c\lambda}}{\lambda}d\lambda$ for constant $c$, we expand the
exponential function $f(\lambda)=e^{-c\lambda}$ using a Taylor
series. The Taylor series at the point $\hat{\lambda}$ is
\begin{equation}\label{eq:tay}
\tilde{t}(\lambda)=\sum_{n=0}^{\infty}\frac{(-c)^{n}e^{-c\hat{\lambda}}}{n!}(\lambda-\hat{\lambda})^n.
\end{equation}
With the help of (\ref{eq:tay}), the term
$\int_{}^{}\frac{e^{-c\lambda}}{\lambda}d\lambda$ in (\ref{eq:power1})
can be further expressed as
\begin{eqnarray}
g_{1}(c,\lambda)&=&\int_{}^{}\frac{e^{-c\lambda}}{\lambda}d\lambda=\int_{}^{}\frac{\tilde{t}(\lambda)}{\lambda}d\lambda\nonumber\\
&=&\int_{}^{}\sum_{n=0}^{\infty}\frac{(-c)^{n}e^{-c\hat{\lambda}}}{n!}\sum_{k=0}^{n}\binom{n}{k}(-\hat{\lambda})^{(n-k)}\lambda^{k-1} d\lambda\nonumber\\
&=&\int_{}^{}\frac{e^{-c\hat{\lambda}}}{\lambda}+\sum_{n=1}^{\infty}\frac{(-c)^{n}e^{-c\hat{\lambda}}}{n!}\left(\frac{(-\hat{\lambda})^n}{\lambda}+\sum_{k=1}^{n}\binom{n}{k}(-\hat{\lambda})^{(n-k)}\lambda^{k-1}\right) d\lambda\nonumber\\
&=&e^{-c\hat\lambda}\ln\lambda+\sum_{n=1}^{\infty}\frac{(-c)^{n}e^{-c\hat{\lambda}}}{n!}\left((-\hat{\lambda})^{n}\ln\lambda+\sum_{k=1}^{n}\binom{n}{k}(-\hat{\lambda})^{(n-k)}\frac{\lambda^{k}}{k}\right),
\end{eqnarray}
and the average transmit power for link $k$ can be written using the
following expression\footnote{For $g_{1}(c,\lambda)$, 
faster convergence of the series can be obtained if $\hat{\lambda}$ is selected as
$\hat{\lambda}=\frac{\lambda_{\max}}{2}$.}:
\begin{eqnarray}\label{eq:twobound}
\mathbb{E}\{P_{k}^{M=2}\}&=& SINR_{t}d_{kk}^\alpha\sigma^2\big((1-\lambda_{\max})e^{-\lambda_{\max}}-(1-\lambda_{\min,k})e^{-\lambda_{\min,k}}+2g_{1}(1,\lambda_{\max})\nonumber\\
&&-2g_{1}(1,\lambda_{\min,k})-g_{1}(2,\lambda_{\max})+g_{1}(2,\lambda_{\min,k})\big).
\end{eqnarray}

Similarly, when $M=4$, the PDF of $\lambda_{1}$ can be expressed as: 
\begin{equation}{\label{eq:pdffour}}
f_{\lambda_{1}}^{M=4}(\lambda)=f_{1}^{'}(\lambda)+f_{2}^{'}(\lambda)+f_{3}^{'}(\lambda)+f_{4}^{'}(\lambda),
\end{equation}
where the definitions of $f_{1}^{'}(\lambda)$, $f_{1}^{'}(\lambda)$,
$f_{1}^{'}(\lambda)$, $f_{1}^{'}(\lambda)$ can be found in the
appendix.  The average transmit power for link $k$ in this case is given by
\begin{eqnarray}
\mathbb{E}\{P_{k}^{M=4}\} & = & \int_{\lambda_{\min,k}}^{\lambda_{\max}}\frac{SINR_{t}d_{kk}^\alpha\sigma^2}{\lambda}f_{\lambda_1}^{M=4}(\lambda)d\lambda \nonumber\\
&=& SINR_{t}d_{kk}^\alpha\sigma^2\int_{\lambda_{\min,k}}^{\lambda_{\max}}\frac{1}{\lambda}\Big(f_{1}^{'}(\lambda)+f_{2}^{'}(\lambda)+f_{3}^{'}(\lambda)+f_{4}^{'}(\lambda)\Big)d\lambda \; . {\label{eq:fourpower}}
\end{eqnarray}
Based on (\ref{eq:fourone})-(\ref{eq:fourfour}) in the appendix, the average transmit power in (\ref{eq:fourpower}) can be evaluated as
\begin{eqnarray}\label{eq:fourbound}
\mathbb{E}\{P_{k}^{M=4}\}&=&SINR_{t}d_{kk}^{\alpha}\sigma^2\Big(\hat{f}_{1}(\lambda_{\max})-\hat{f}_{1}(\lambda_{\min,k})+\hat{f}_{2}(\lambda_{\max})-\hat{f}_{2}(\lambda_{\min,k})+\hat{f}_{3}(\lambda_{\max})\nonumber\\
&&-\hat{f}_{3}(\lambda_{\min,k})+\hat{f}_{4}(\lambda_{\max})-\hat{f}_{4}(\lambda_{\min,k})\Big).
\end{eqnarray}

For the single-antenna case, we average the transmit power over the channel 
gain $h_{kk}$ to determine the average transmit power of link $k$ as
\begin{eqnarray}
\mathbb{E}\{P_{k}^{M=1}\} & = & \int_{\sqrt{\lambda_{\min,k}}}^{\lambda_{\max}}\frac{2SINR_{t}d_{kk}^{\alpha}\sigma^2}{h}e^{-h^{2}} dh \nonumber\\
&=& 2SINR_{t}d_{kk}^{\alpha}\sigma^2\Big(\tilde{s}(\lambda_{\max})-\tilde{s}\big(\sqrt{\lambda_{\min,k}}\big)\Big),
\end{eqnarray}
where the function $\tilde{s}(x)$ is defined as
\begin{equation}
\tilde{s}(x)=\int\frac{e^{-x^2}}{x}dx=\ln{x}+\sum_{n=1}^{\infty}(-1)^{n}\frac{x^{2n}}{2n\cdot n!}.
\end{equation}
Thus, the lower bound on the average sum transmit power of the $N_{L}$ links is given by
\begin{equation}
P_{B}^{M}=\frac{1}{N_s}\sum_{k=1}^{N_{L}}\mathbb{E}\{P_{k}^{M}\}.
\end{equation}
\section{Simulation Results}\label{sec:sim}

\begin{figure}
\begin{center}
\includegraphics[height=3.5in, width=4.8in]{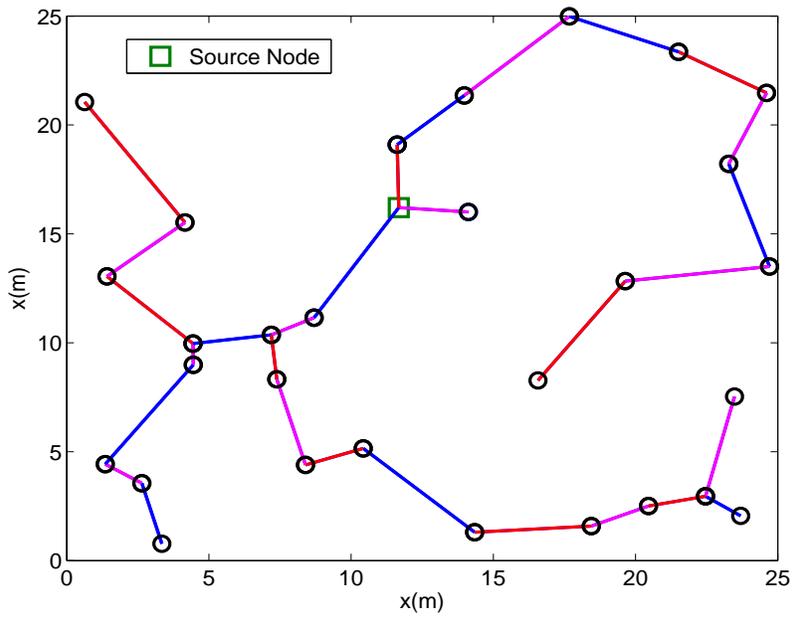}
\caption{The topology of the network assumed for the simulation with $N=30$ nodes.}\label{f_3}
\end{center}
\end{figure}

For our simulations, we consider a network with $N=30$ nodes,
uniformly distributed in a $25\textrm{m}\times 25\textrm{m}$ area as
shown in Fig.~\ref{f_3}. The node represented by a square is the
source node, and the edges represent the $N_L=29$ links in the
multicast tree. Also in Fig.~\ref{f_3}, the
scheduling result for a single channel realization is provided when
$N_s=3$, $M=4$, $\gamma=0.04$ and $C_{req}=0.9\textrm{bps/Hz}$; the links that have the same color have been scheduled to
transmit in the same time slot. We assume noise with unit power, a maximum transmit
power of $P_{\max}=25\textrm{dB}$ for each node, and a path loss
exponent of $\alpha=2$. The connectivity performance and the sum
transmit power of the network are averaged over 300 independent
channel realizations and the performance for different $M$, $N_{s}$
and $\gamma$ is provided. Note that although the highest collision
degree for any of the nodes is 4, the minimum number of slots
considered for a given frame is 3, which means our scheduling
algorithm can use only 3 time slots to completely avoid the link
collisions.

\begin{figure}
\centering
\includegraphics[height=3.5in, width=4.8in]{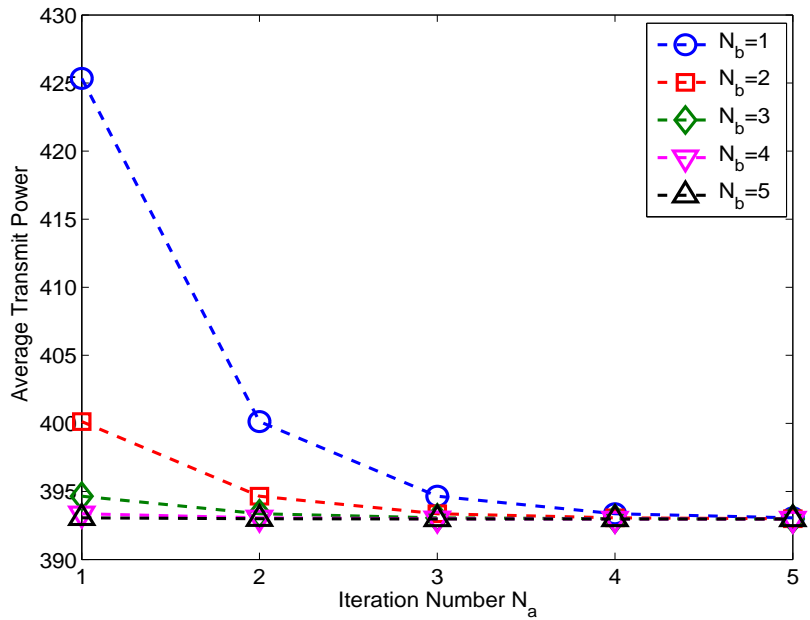}
\caption{Average sum transmit power comparison between different $N_a$ and $N_b$, with $C_{req}=0.9\text{bps/Hz}$, $\gamma=0.01$, $N_s=5$ for $M=4$.}\label{f_4}
\end{figure}

In the power control algorithm, only the first
few iterations play an important role in the algorithm performance.
This is illustrated in the example of Fig.~\ref{f_4} for a case with
$M=4$, $C_{req}=0.9$ bps/Hz, $\gamma=0.1$ and $N_s=5$.  We see that
all of the power reduction occurs for $\{N_a,N_b\} \le 3$.  In the
simulation results that follow, we set $N_a=3$ and $N_b=2$.  The
minimum rate requirement $C_{req}$ in the following simulations is
assumed to be adjusted to take into account the overhead due to
channel estimation and feedback to the source. The QoS requirements are set to be $C_{req}=0.9\text{bps/Hz}$ for
$M=4$, $C_{req}=0.5\text{bps/Hz}$ for $M=2$ and $C_{req}=0.1\text{bps/Hz}$
for $M=1$, and plots for both $\gamma=0.01$ and $\gamma=0.04$ are included.
We show results for different $C_{req}$ with each $M$, due to the strong impact of the
number of antennas on performance. When the
connectivity probability is near 1, the approximate total throughput
of the network can be found by multiplying $C_{req}$ by $N_L-1$ and
the connectivity probability. 

We compare the performance of three
cases: outdated CSI (OCSI), local perfect CSI (LPCSI) and global
perfect CSI (GPCSI). For OCSI, the source node uses the outdated
global CSI $\bar{\mathcal{H}}$ as the input to the scheduling
algorithm, and the links transmit according to the scheduling results
$\mathcal{S}$, $\mathcal{T}$ and $\mathcal{P}$ provided by the source.
In LPCSI, the CSI of the links transmitting in the same time slot are
assumed to be known by the other active links and, based
on this local CSI, the transmit beamforming algorithm is used to
re-optimize the beamformers for that time slot.  The performance gain
of LPCSI over OCSI represents the advantage provided by the use of local
instantaneous CSI. For GPCSI, we assume the source node has perfect
knowledge of $\mathcal{H}$, which amounts to assuming $\gamma=0$.  For
this case, the lower bound on SINR in (\ref{eq:opt4}) is replaced with
the exact SINR expression in~(\ref{eq:sinra}). The results for GPCSI
indicate the best performance that the scheduling algorithm can
achieve, and we see that as predicted, its performance matches the
bound in Section~\ref{sec:conbound}, provided that $N_s$ is chosen
large enough so that the interference can be properly mitigated.

\begin{figure}
\centering
\includegraphics[height=3.3in, width=7in]{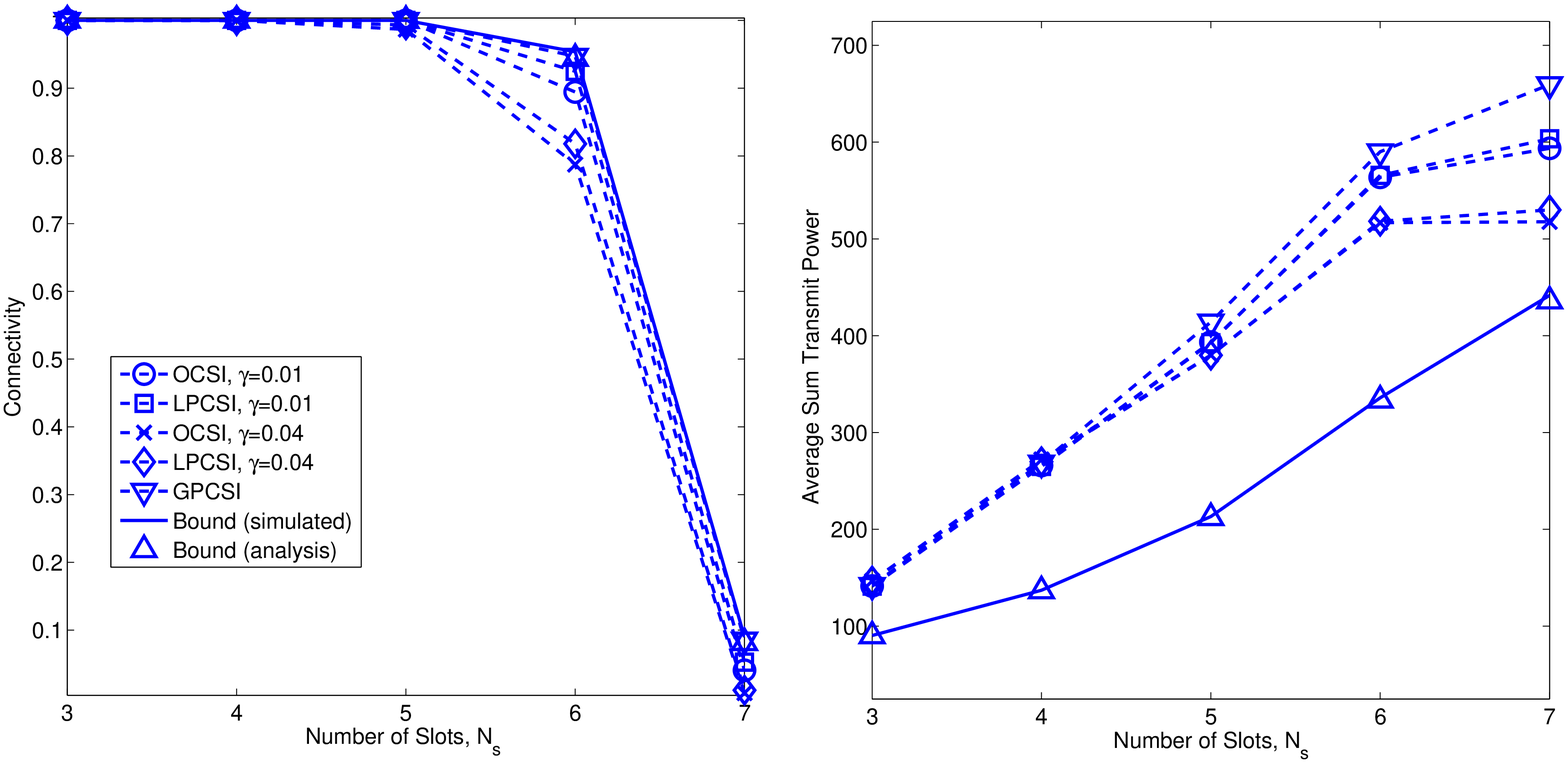}
\caption{Connectivity performance and average sum transmit power of the network for different number of time slots per frame, with $M=4$, $C_{req}=0.9\text{bps/Hz}$.}\label{f_5}
\end{figure}

Fig.~\ref{f_5} provides the connectivity performance and the
average sum transmit power for $M=4$. The behavior of the connectivity metric can be explained as follows.
As the number of colors (slots) increases, the number of links that
transmit simultaneously will be reduced, the interference between the
links will thus be reduced, and the SINR at the receiver of each
link will be improved.  However, to guarantee the spectral efficiency
$C_{req}$, the threshold $SINR_{t}$ must also increase due to the shorter time
slot.  When the benefit brought by the increase in the number of slots
is larger than the penalty caused by the increased $SINR_t$, the
connectivity of the network will increase; otherwise, the connectivity
will decrease.  This trade-off results in an optimal value for $N_s$
for each case considered.

The behavior of the transmit power curves is slightly different, since
links that cannot meet the desired SINR are not allowed to transmit.
This obviously reduces the connectivity metric, but it also reduces
the total transmit power.  That explains why, for example, the
transmit power required by GPCSI is always higher than that for the
other cases; since it achieves a higher connectivity, more links are
active and more transmit power is consumed. Note also that the analytical values
for both the connectivity and transmit power bounds match those obtained
in the simulation. It can be observed that when $\gamma=0.04$, full connectivity is achieved at $C_{req}=0.9\text{bps/Hz}$ with a required transmit power of about 150.
Comparing the performance of GPCSI and OCSI for $\gamma=0.04$, we see no impact of the outdated CSI on the network connectivity for $N_s\le 5$.

\begin{figure}
\centering
\includegraphics[height=3.3in, width=7in]{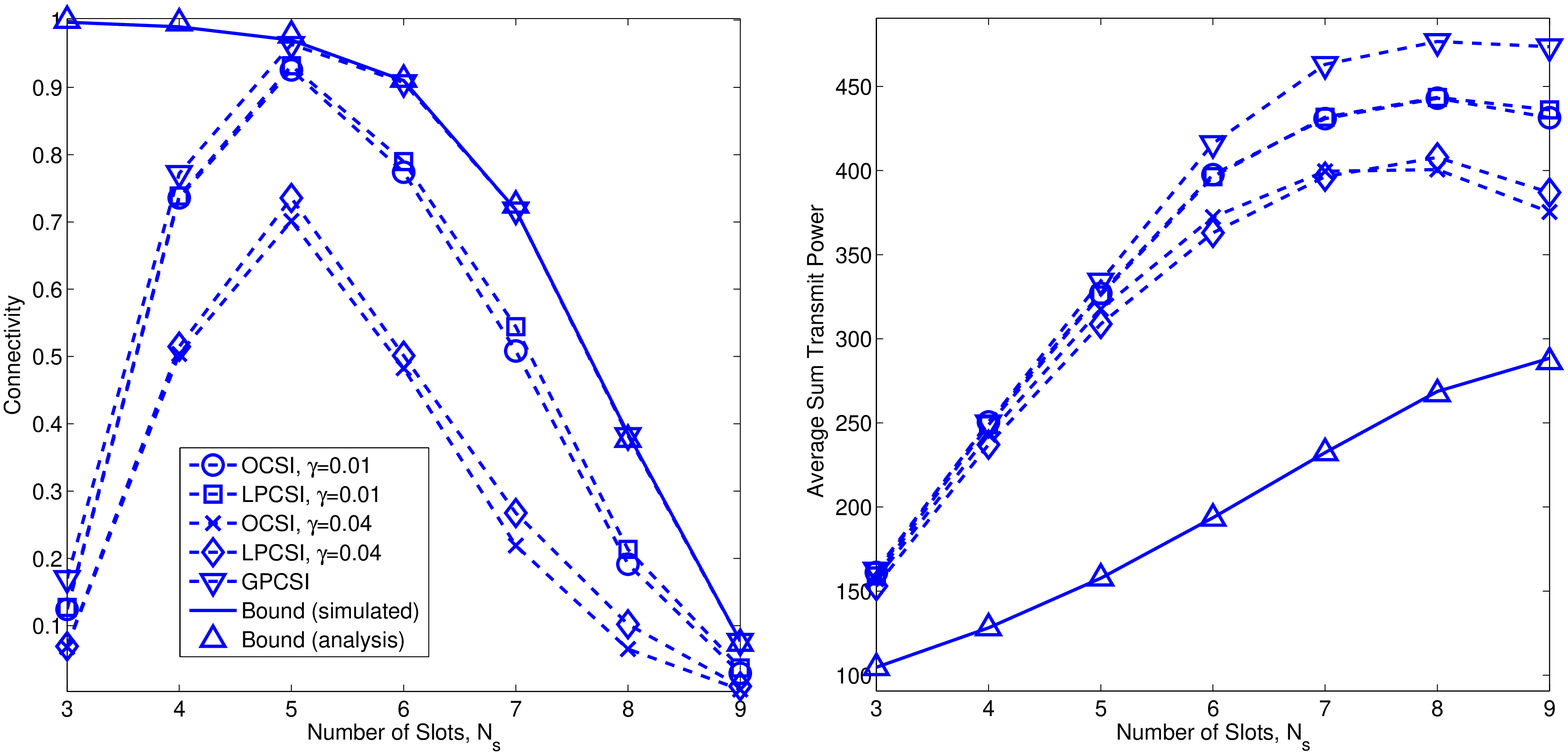}
\caption{Connectivity performance and average sum transmit power of the network for different number of time slots per frame, with $M=2$, $C_{req}=0.5\text{bps/Hz}$.}\label{f_6}
\end{figure}

In Fig. \ref{f_6}, we see that the two-antenna network is able to
achieve a connectivity of about 0.7 at $C_{req}=0.5\text{bps/Hz}$, 
requiring a transmit power of approximately 300. Compared with
GPCSI for $\gamma=0.04$ and $N_s=5$, the network connectivity of OCSI is
reduced by about 25\%.

\begin{figure}
\centering
\includegraphics[height=3.3in, width=7in]{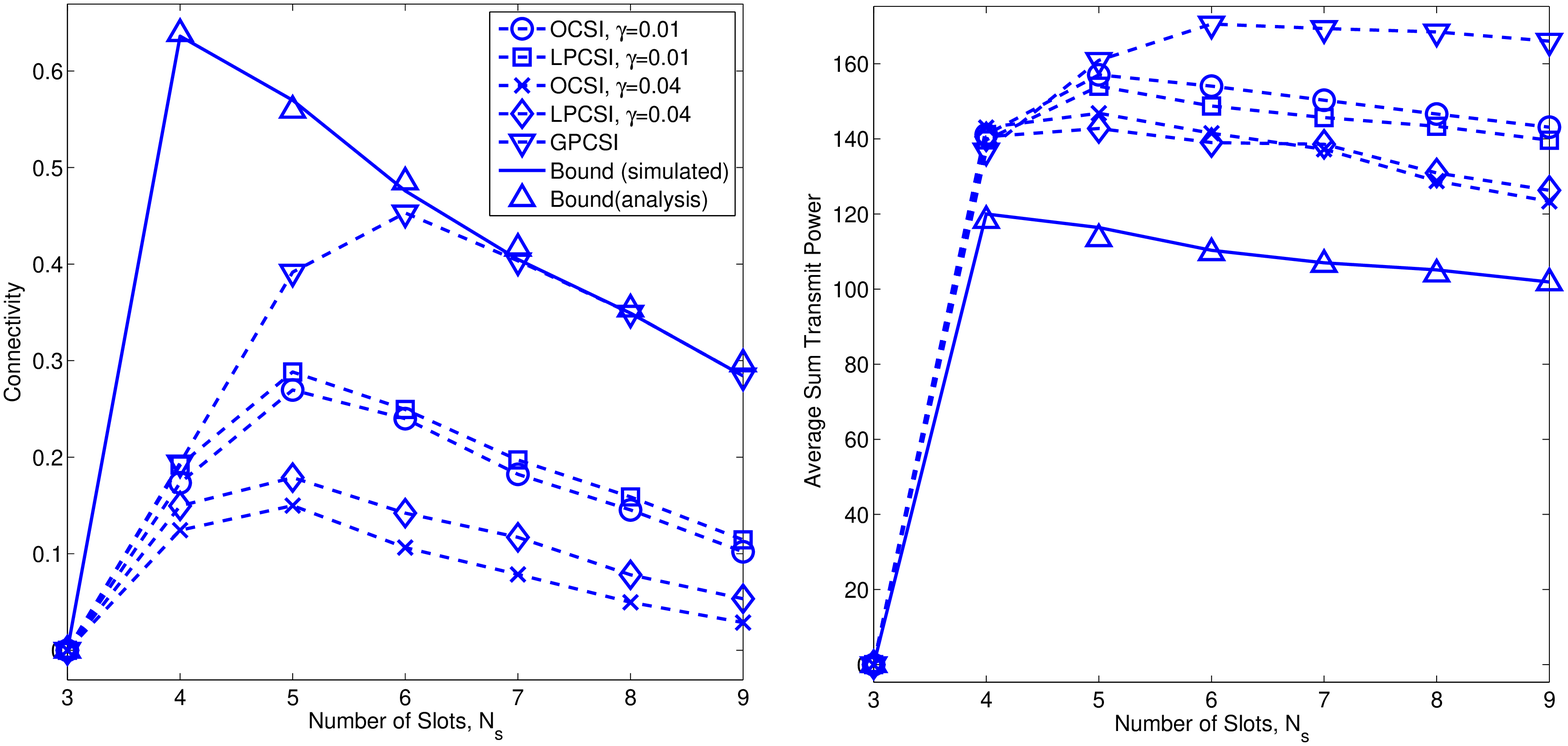}
\caption{Connectivity performance and average sum transmit power of the network for different number of time slots per frame, with $M=1$, $C_{req}=0.1\text{bps/Hz}$.}\label{f_7}
\end{figure}

Fig. \ref{f_7} presents the performance results for $M=1$ with
$C_{req}=0.1\text{bps/Hz}$. Peak connectivity occurs at $N_s=5$, with
a required transmit power of about 140, and a connectivity of only
about 0.15 is achieved. When $\gamma=0.04$, we see that the
connectivity performance of OCSI is reduced by a factor of three with
respect to GPCSI.

Comparing the connectivity performance in Figs. \ref{f_5}-\ref{f_7},
it can be observed that the optimal value for $N_s$ decreases with $M$,
even though we have increased the desired throughput with $M$ as well.
The benefit of having nodes equipped with multiple antennas is clearly
evident in terms of connectivity and total network throughput. Note
that the benefit of nodes with multiple antennas also manifests itself
in terms of the transmit power required to achieve maximum
connectivity. Instead of a throughput gain of four with the same level
of reliability, which might be expected when comparing the $M=4$ and
$M=1$ networks, we see that there is a multiplicative benefit that
results from using multiple antennas for improved connectivity. The
additional spatial degrees of freedom reduce interference between
links within the same time slot, they reduce the total number of
required time slots, and they reduce the amount of transmit power (and
hence interference to other links) to achieve a given throughput.  The
combined gain of these effects can only be observed by considering the
joint optimization of the transmit beamformers, scheduling and power
control.

Further evidence of the multiplicative benefit of multiple
antennas is the fact that we observe a decrease in sensitivity to
imprecise CSI as the number of antennas at each node increases.
Note also that using locally accurate CSI to adjust the transmit beamformers (LPCSI) has only a slight
benefit in improving performance relative to using the beamformers
based on outdated CSI (OCSI).  This is due to the fact that only a
single data stream is transmitted on each link.  If multiple data
streams were allowed, the importance of locally accurate CSI would be
more critical; one could potentially improve connectivity, but the
performance gain would be much more sensitive to imprecise CSI.

\begin{figure}
\centering
\includegraphics[height=3.5in, width=7in]{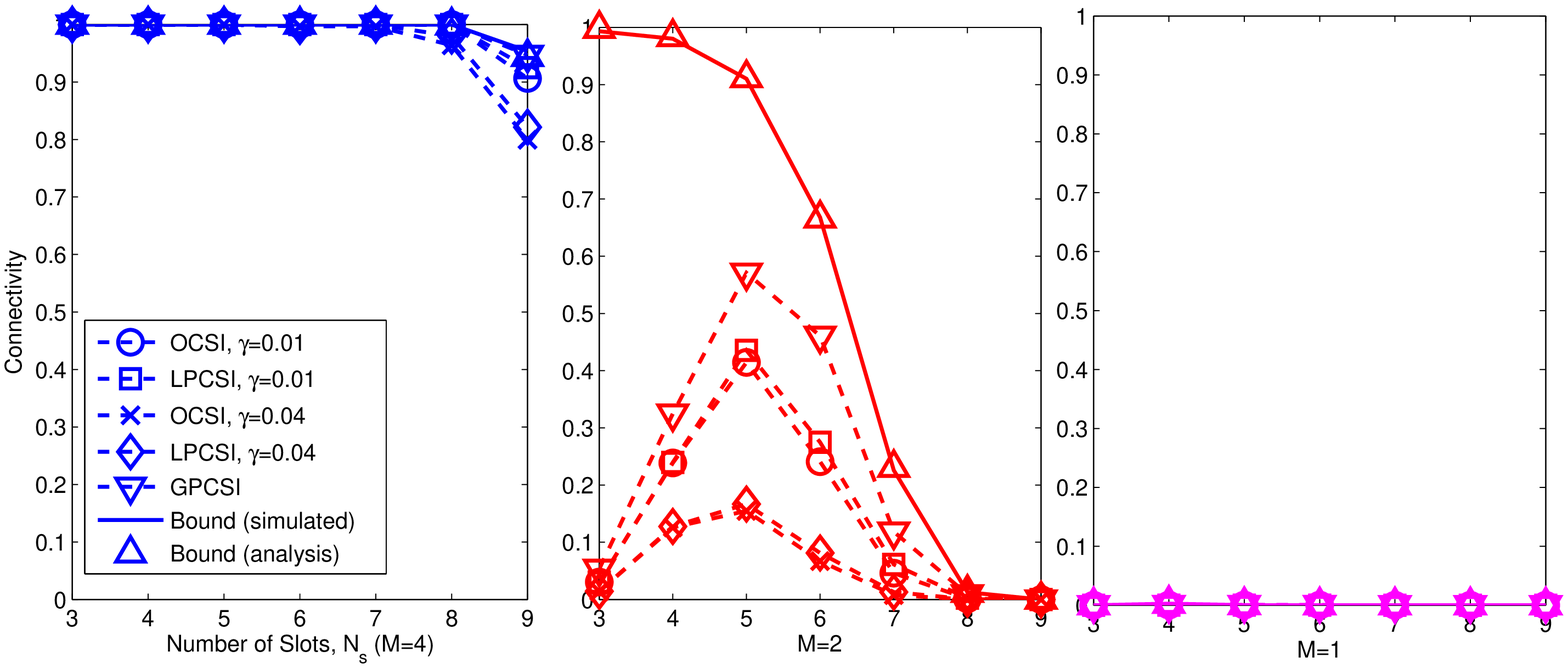}
\caption{Connectivity comparison between different antenna configurations $M=4$, $M=2$, $M=1$, with $C_{req}=0.6\text{bps/Hz}$, $\gamma=0.01$ or $\gamma=0.04$.}\label{f_8}
\end{figure}

In Fig.~\ref{f_8}, we show the connectivity results for all
three antenna sizes for the same value of $C_{req}=0.6$bps/Hz. The
$M=4$ network achieves full connectivity in this case, while the $M=1$
network is nearly disconnected.  The two-antenna case falls somewhere
in between, with performance depending on how accurate the CSI is at
the scheduler. 

\begin{figure}
\centering
\includegraphics[height=3.5in, width=7in]{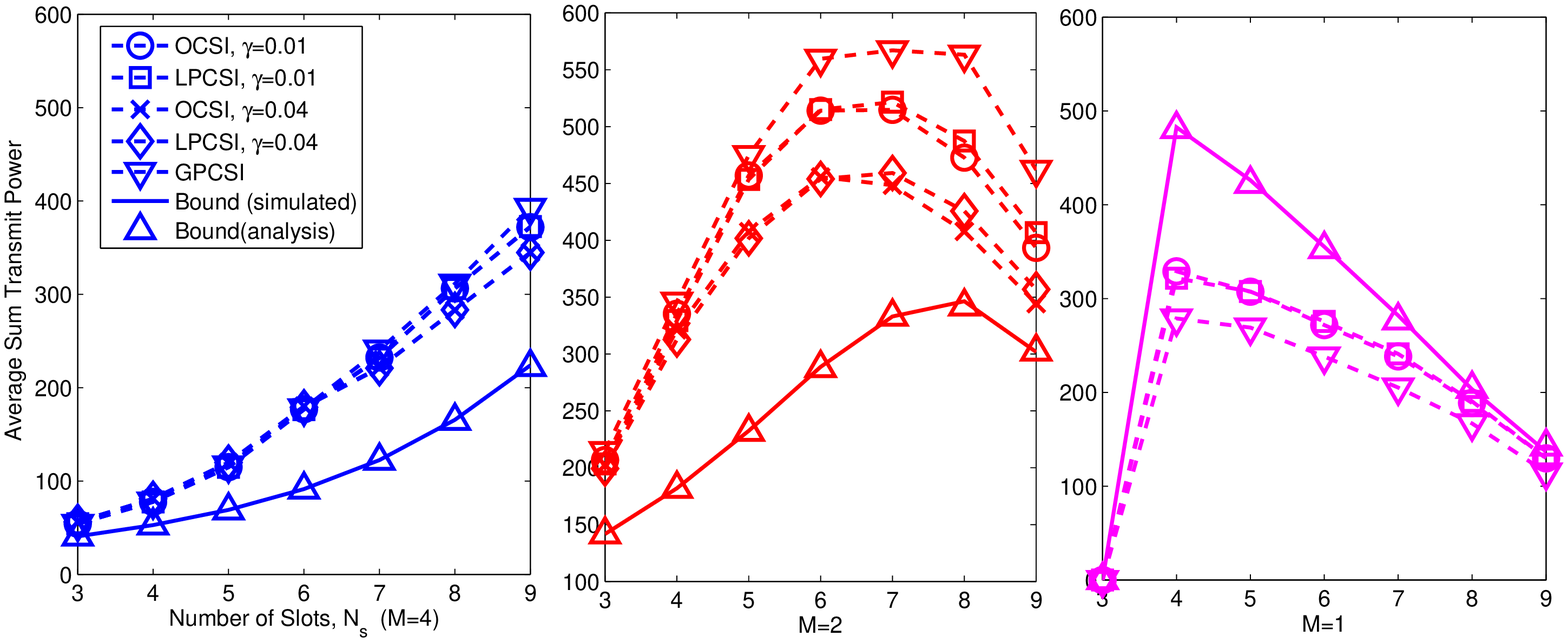}
\caption{Average sum transmit power comparison between different antenna configurations $M=4$, $M=2$, $M=1$, with $C_{req}=0.6\text{bps/Hz}$, $\gamma=0.01$ or $\gamma=0.04$.}\label{f_9}
\end{figure}

In Fig.~\ref{f_9}, the transmit power results are provided. The
difference in transmit power is especially evident in this
example. The $M=4$ network requires a factor of seven times less power
to achieve a connectivity that is five times higher than the $M=2$
case. While the single-antenna network is almost disconnected, the
fact that the average sum transmit power is non-zero indicates that at
least some of the links are active.

\begin{figure}
\centering
\includegraphics[height=3.5in, width=4.8in]{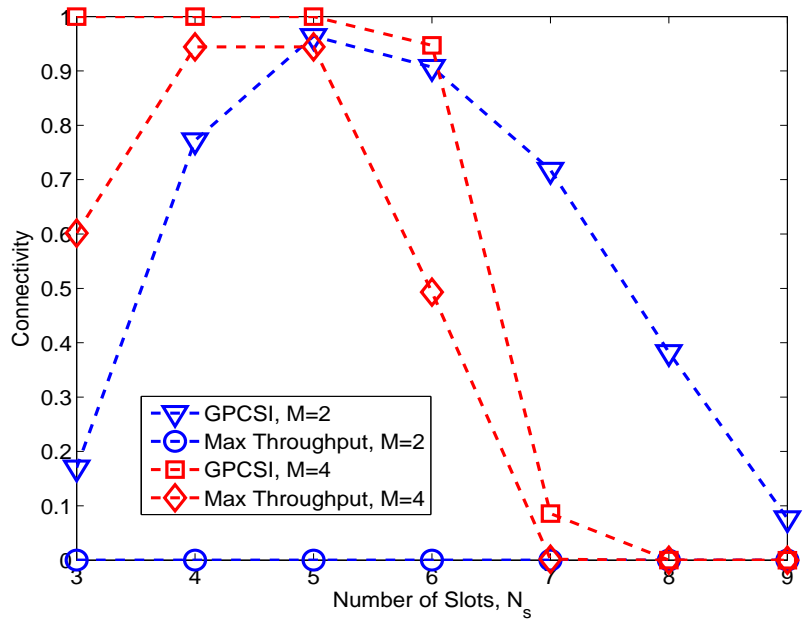}
\caption{Connectivity performance of the network for different scheduling methods, with $C_{req}=0.5\text{bps/Hz}$ for $M=2$ and $C_{req}=0.9\text{bps/Hz}$ for $M=4$.}\label{f_10}
\end{figure}

\begin{figure}
\centering
\includegraphics[height=3.5in, width=4.8in]{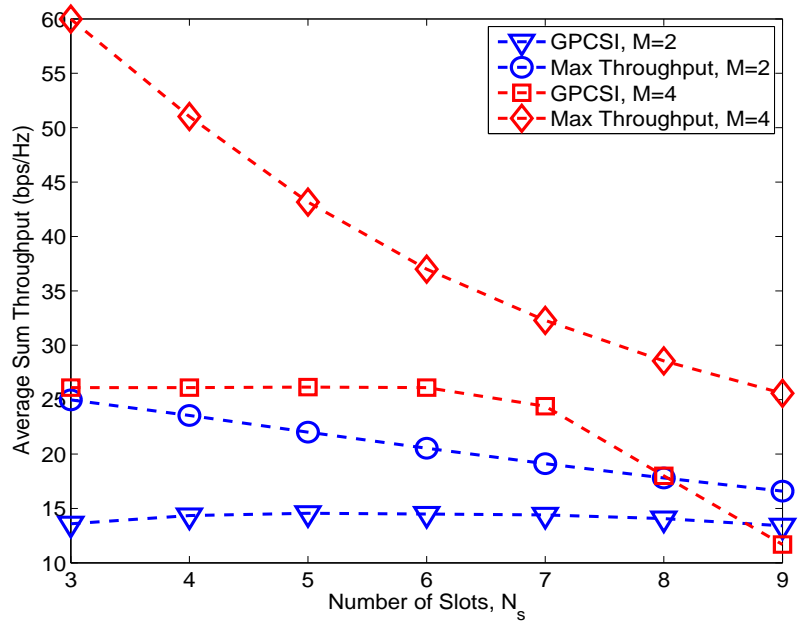}
\caption{Average sum throughput comparison between different scheduling methods, with $C_{req}=0.5\text{bps/Hz}$ for $M=2$ and $C_{req}=0.9\text{bps/Hz}$ for $M=4$.}\label{f_11}
\end{figure}

\begin{figure}
\centering
\includegraphics[height=3.5in, width=4.8in]{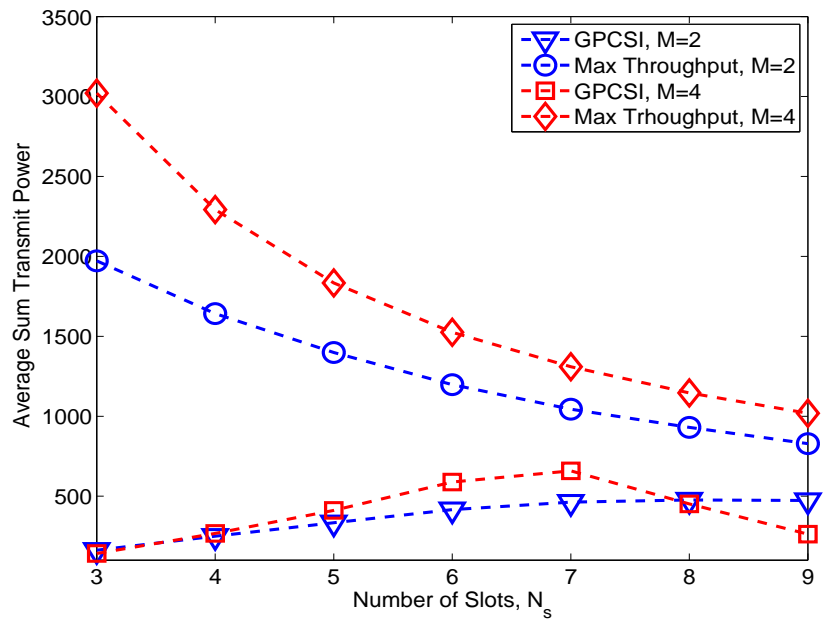}
\caption{Average sum transmit power comparison between different scheduling methods, with $C_{req}=0.5\text{bps/Hz}$ for $M=2$ and $C_{req}=0.9\text{bps/Hz}$ for $M=4$.}\label{f_12}
\end{figure}

Finally, in Figs.~\ref{f_10}-\ref{f_12}, we compare the performance of
our connectivity-based approach with one that attempts to choose the
network parameters in order to maximize the throughput of the network.
The optimization algorithm in this case is similar to the scheduling
algorithm for connectivity, with the exception that the objective
function in (\ref{eq:opt3}) is replaced with
$\sum_{k=1}^{N_L}\log_2(1+SINR_k)$.  In Fig.~\ref{f_10}, for $M=2$, we
see that attempting to maximize the throughput results in zero
connectivity for the network, while using our approach, we achieve a
connectivity in excess of 0.95 for the optimal value of $N_s=5$.

In Fig.~\ref{f_11}, the overall network throughput of the
max-throughput approach is about 50\% higher than the connectivity
optimization method.  This indicates that a subset of the nodes is
able to communicate at a higher rate, but the network is disconnected
and the message from the source is not reaching all of the network
nodes.

In Fig.~\ref{f_12}, the results show that the proposed approach
achieves the maximum connectivity with about five times less transmit
power than the max-throughput approach. The connectivity of the
max-throughput approach improves when $M=4$, but with an increase in
required power consumption of nearly a factor of eight.

\section{Conclusions}\label{sec:conc}

In this paper we investigated the use and benefit of multiple antennas
in an ad hoc network with a source streaming data to all nodes via a
multi-hop tree.  Based on a novel definition of network connectivity,
a scheduling algorithm was developed that takes advantage of the
interference mitigation capabilities of the MIMO nodes.  A key
component of the algorithm is the design of transmit beamformers for
simultaneously active nodes that optimizes the connectivity metric.
Ultimately, the scheduling algorithm acts to break down the full
network into a set of smaller interference networks whose links 
are able to be simultaneously active due to the interference mitigation
provided by the multiple antennas.  We also derived performance bounds
on the network connectivity and average required sum transmit power,
assuming zero interference and perfect CSI.  These bounds represent
ultimate limits for the performance of the network, and were used in
the simulations to compare against the actual behavior of the network.
Our simulation results indicate the significant advantage provided
by multiple antennas in the ad hoc network, in terms of connectivity,
throughput, reduced transmit power and resilience to outdated CSI.  
By considering the joint problem of transmit beamformer design,
scheduling and power control, we observe a multiplicative benefit to
the use of multiple antennas for improving network reliability.  Such
an observation could not be made by solving each of these problems in
isolation from the others.



\appendix
The functions $f_{1}(\lambda)$, $f_{2}(\lambda)$, $f_{3}(\lambda)$, $f_{4}(\lambda)$ in (\ref{eq:cdffour}) are defined as follows:
\begin{eqnarray}
f_{1}(\lambda)&=&(e^{-\lambda}+e^{-3\lambda})\left(6\lambda^2+\frac{11}{6}\lambda^4+\frac{1}{36}\lambda^6\right),\nonumber\\
f_{2}(\lambda)&=&(e^{-\lambda}-e^{-3\lambda})\left(-4\lambda^3-\frac{1}{3}\lambda^5\right),\nonumber\\
f_{3}(\lambda)&=&-e^{-2\lambda}\left(12\lambda^2+\frac{2}{3}\lambda^4+\frac{2}{9}\lambda^6+\frac{1}{144}\lambda^8\right),\nonumber\\
f_{4}(\lambda)&=&4e^{-\lambda}+4e^{-3\lambda}-6e^{-2\lambda}-e^{-4\lambda}\nonumber.
\end{eqnarray}
Based on $\!f_{1}(\lambda)$, $\!f_{2}(\lambda)$, $\!f_{3}(\lambda)$, $\!f_{4}(\lambda)$, the functions $\!f_{1}^{'}(\lambda)$, $\!f_{2}^{'}(\lambda)$, $\!f_{3}^{'}(\lambda)$, $\!f_{4}^{'}(\lambda)$ in (\ref{eq:pdffour}) are defined as:
\begin{eqnarray}
f_{1}^{'}(\lambda)&=&-\frac{d f_{1}(\lambda)}{d \lambda}=(e^{-\lambda}+e^{-3\lambda})\left(-12\lambda-\frac{22}{3}\lambda^{3}-\frac{1}{6}\lambda^{5}\right)-(e^{-\lambda}+3e^{-3\lambda})\left(-6\lambda^{2}-\frac{11}{6}\lambda^{4}-\frac{1}{36}\lambda^{6}\right),\nonumber\\
f_{2}^{'}(\lambda)&=&-\frac{d f_{2}(\lambda)}{d \lambda}=(-e^{-\lambda}+3e^{-3\lambda})\left(4\lambda^{3}+\frac{1}{3}\lambda^{5}\right)+(e^{-\lambda}-e^{-3\lambda})\left(12\lambda^{2}+\frac{5}{3}\lambda^{4}\right),\nonumber\\
f_{3}^{'}(\lambda)&=&-\frac{d f_{3}(\lambda)}{d \lambda}=e^{-2\lambda}\left(24\lambda+\frac{8}{3}\lambda^{3}+\frac{4}{3}\lambda^{5}+\frac{1}{18}\lambda^{7}\right)-2e^{-2\lambda}\left(12\lambda^2+\frac{2}{3}\lambda^4+\frac{2}{9}\lambda^6+\frac{1}{144}\lambda^8\right),\nonumber\\
f_{4}^{'}(\lambda)&=&-\frac{d f_{4}(\lambda)}{d \lambda}=4e^{-\lambda}+12e^{-3\lambda}-12e^{-2\lambda}-4e^{-4\lambda}\nonumber.
\end{eqnarray}

Define $g_{2}(c,n,x)=\int_{}^{}x^{n}e^{cx}dx=\sum_{i=0}^{n}\frac{n!}{c^{i+1}(n-i)!}x^{n-i}e^{cx}$, then it can be shown that
\begin{align}{\label{eq:fourone}}
\hskip 2.3em\hat{f}_{1}(\lambda)=&\int\frac{f_{1}^{'}(\lambda)}{\lambda}d\lambda\nonumber\\
=&-12\Big(g_{2}(-1,0,\lambda)\!+\!g_{2}(-3,0,\lambda)\Big)\!-\!\frac{22}{3}\Big(g_{2}(-1,2,\lambda)\!+\!g_{2}(-3,2,\lambda)\Big)\nonumber\\
&-\!\frac{1}{6}\Big(g_{2}(-1,4,\lambda)\!+\!g_{2}(-3,4,\lambda)\Big)+6\Big(g_{2}(-1,1,\lambda)\!+\!3g_{2}(-3,1,\lambda)\Big)\nonumber\\
&+\frac{11}{6}\Big(g_{2}(-1,3,\lambda)+3g_{2}(-3,3,\lambda)\Big)+\frac{1}{36}\Big(g_{2}(-1,5,\lambda)+3g_{2}(-3,5,\lambda)\Big),
\end{align}
\begin{align}{\label{eq:fourtwo}}
\hskip -9.5em\hat{f}_{2}(\lambda)=&\int_{}^{}\frac{f_{2}^{'}(\lambda)}{\lambda}d\lambda\nonumber\\
=&4\Big(3g_{2}(-3,2,\lambda)\!-\!g_{2}(-1,2,\lambda)\Big)\!+\!\frac{1}{3}\Big(3g_{2}(-3,4,\lambda)\!-\!g_{2}(-1,4,\lambda)\Big)\nonumber\\
&+\!12\Big(g_{2}(-1,1,\lambda)\!-\!g_{2}(-3,1,\lambda)\Big)+\!\frac{5}{3}\Big(g_{2}(-1,3,\lambda)\!-\!g_{2}(-3,3,\lambda)\Big),
\end{align}
\begin{align}{\label{eq:fourthree}}
\hskip -7.5em\hat{f}_{3}(\lambda)=&\int_{}^{}\frac{f_{3}^{'}(\lambda)}{\lambda}d\lambda\nonumber\\
=&24g_{2}(-2,0,\lambda)+\frac{8}{3}g_{2}(-2,2,\lambda)+\frac{4}{3}g_{2}(-2,4,\lambda)+\frac{1}{18}g_{2}(-2,6,\lambda)\nonumber\\
&-24g_{2}(-2,1,\lambda)-\frac{4}{3}g_{2}(-2,3,\lambda)-\frac{4}{9}g_{2}(-2,5,\lambda)-\frac{1}{72}g_{2}(-2,7,\lambda),
\end{align}
\begin{eqnarray}{\label{eq:fourfour}}
\hskip -4em\hat{f}_{4}(\lambda)=\int_{}^{}\frac{f_{4}^{'}(\lambda)}{\lambda}d\lambda=4\big(g_{1}(1,\lambda)-g_{1}(4,\lambda)\big)+12\big(g_{1}(3,\lambda)-g_{1}(2,\lambda)\big).
\end{eqnarray}



%
\bibliography{reference}

\newpage

\end{document}